\documentclass[aps,pre,twocolumn,showpacs,10pt]{revtex4-1}
\usepackage{color}
\usepackage{amsmath}
\usepackage{amsfonts}
\usepackage{amssymb}
\usepackage{graphicx}

\begin{document}
\title{Phase Diagram of a Reentrant Gel of Patchy Particles}

\author{S{\'andalo} Rold{\'a}n-Vargas$^1$, 
Frank Smallenburg$^1$, 
Walter Kob$^2$, 
Francesco Sciortino$^1$
}
\affiliation{ $^1$Department of Physics, Sapienza, Universit\`{a}
di Roma, Piazzale Aldo Moro 2, I-00185, Roma, Italy,\\ $^2$Laboratoire
Charles Coulomb, UMR 5221, CNRS and Universit\'e Montpellier 2,
34095 Montpellier, France}

\begin{abstract}
We study the phase diagram of a binary mixture of patchy particles
which has been designed to form a reversible gel. For this we perform
Monte Carlo and molecular dynamics simulations to investigate the
thermodynamics of such a system and compare our numerical results with
predictions based on the analytical parameter-free Wertheim theory. We
explore a wide range of the temperature-density-composition space that
defines the three-dimensional phase diagram of the system. As a result,
we delimit the  region of thermodynamic stability of the fluid. We find
that for a large region of the phase diagram the Wertheim theory is
able to give a quantitative description of the system. For higher
densities our simulations show that the system is crystallizing into a
BCC structure. Finally we study the relaxation dynamics of the system
by means of the density and temperature dependences of the diffusion
coefficient. We show that there exists a density range where the system
passes reversibly from a gel to a fluid upon both heating and cooling,
encountering neither demixing nor phase separation.
\end{abstract}

\maketitle

\section{INTRODUCTION}
In recent times we have observed a very fruitful symbiosis in the
field of soft matter physics: The interplay between the synthesis of
nano-and meso-sized particles with a highly versatile functionality
\cite{dnapatchy1,newsandviews,granickpatchy2011,laponite,SEEMAN_03,CONDON_06,dnatetramersJACS,dnabellini,kegelpnas,granickbook,janussoftmatter,granickacs,iwashita}
and the computational design of new model systems
\cite{glotzer2007,dijkstrapolyhedra,romano_on_granick,frenkelnatmat,smallenburg2012vacancy,likosbianchi,zdeneck,cacciuto,likosstar}.
This combined effort can favor  the emergence of new materials
which have  the potential for novel
applications. It can also allow us to obtain a deeper understanding
of some intriguing states of matter such as glasses or gels
\cite{zaccareview,recentreview}. Thus the old problem of statistical
physics, to connect the macroscopic properties of a system with its
\textit{immutable} microscopic constituents, has now been extended with
the possibility of an \textit{ad hoc} design of the today controllable
primary particles.

This bottom-up design can capitalize on the idea of
\textit{competitive interactions}, i.e. the fact that the
competition between distinct interaction mechanisms can cause
the system to self-assemble into  different local structures
\cite{sear,kegelcluster,mossa,safran,russoj}. For example, the
interactions between  particles can be designed in such a way that at
intermediate temperatures the system presents a structure favoured by
entropy whereas at low temperature it has a structure that
is stabilized by potential energy \cite{frenkelnatmat,inversegel}.

Due to their directional and selective interactions, patchy colloidal
particles \cite{newsandviews,dnapatchy1,glad1,flavioNC} have been found
to allow a precise control of these competitive interaction mechanisms
\cite{russoj,inversegel}. Recently explored examples for this include
the competition between chaining and branching in patchy colloids,
where a specific design of the inter-patch interactions results
in a phase diagram in which the density of the coexisting liquid
approaches the density of the gas \cite{russoj}. Another example is
given by colloids that are coated with two different DNA sequences,
establishing a competition between intra- and inter-particle interactions,
and providing a way to tune the effective inter-particle interaction to
favor crystal formation \cite{frenkelnatmat}.  Patchy particles can also
be designed to induce, via  a subtle entropy/enthalpy compensation
mechanism, closed-loop phase diagrams where the low-temperature
stable phase is a fluid of self-assembled weakly interacting clusters
\cite{prljanus,januspccp,doyejcp,lorenzoring,almarzaring}.

In the present work we explore one particular example of competing
interactions by studying in detail the phase diagram of a binary
mixture of patchy particles that has been specifically designed to
create a material that continuously and reversibly passes from a fluid
to a gel and from the gel to a fluid upon cooling \cite{inversegel}.
The first species of particles, $A$, in this binary mixture consists
of particles with valence (functionality) four (i.e. particles with four attractive
patches) that at intermediate temperature form an entropically favorable
random tetrahedral network (the gel) structurally similar to that present
in atomistic systems such as silica or silicon \cite{binder_11,saika2013understanding}. Upon a
further decrease in temperature this network is fragmented by the second
species ($B$) of mono-valence particles which compete for bonding with the
patches of species $A$. The possibility of fragmenting the $AA$-gel is
related to the fact that if a sufficiently large number of $B$-particles
is available, each $A$-particle can form up to four $AB$-bonds, thus
leading to a energetically favorable structure.  In contrast, in the fully
bonded network of $A$-particles, the number of $AA$-bonds is only two
per $A$-particle. Thus, with appropriate choices for the composition
of the system and the relative strengths of two types of bonds, it
can be  energetically more favorable to form $AB$-bonds. As a result,
the $B$-particles progressively block the network connectivity as the
temperature decreases, thus returning the system to the fluid state.

In contrast to our first study (Ref.~\cite{inversegel}) which focused
on one fixed density and one fixed composition, here we investigate
the full temperature-density-composition space of the model by Monte
Carlo and event driven molecular dynamics simulations and compare these
results with the ones obtained in the framework of Wertheim's theory
\cite{wertheim1,wertheim2}. The use of these techniques allows us to
calculate with precision the coexistence region of our system and thus to
determine the range of densities and compositions at which the reversible
gel is thermodynamically stable. We find that at low densities the system
undergoes a phase separation into an $A$-rich network phase and a $B$-rich
gas phase, while at high densities the system crystallizes. Finally,
our study provides one of the first tests for verifying the predictions
of Wertheim's theory for binary mixtures of patchy particles, thus
allowing to discuss the limits of validity of this analytical approach.

The article is organized as follows. In Section II we describe our model
and provide the reader with a brief review on the relevant theoretical
and methodological aspects. In Section III.A we present the study of the
complete three-dimensional phase diagram of our system as obtained from
Wertheim's theory and from our grand canonical Monte Carlo simulations by
using an extension of the successive umbrella sampling for binary mixtures
\cite{lorenzo-sus-2d}. Section III.B concentrates on the study of the
two-dimensional temperature-density cuts of the phase diagram
corresponding to the stoichiometric molar fraction, i.e. the  molar
composition  for which there are four $B$-particles for each $A$-particle.
In Section III.C we document the spontaneous crystallization of our
system --- at the stoichiometric molar fraction --- in molecular dynamics
simulations at high densities and intermediate temperatures.  Also for
this particular composition, in Section III.D we focus our attention on
the dynamics through the study of the temperature and density dependences
of the diffusion coefficient of the $A$-particles. Finally, in Section
IV we conclude with a summary of our main findings.

\section{MODEL AND THEORETICAL BACKGROUND}
\subsection{INTERACTION POTENTIAL: KERN-FRENKEL MODEL}
The particles investigated in this paper are given by the well-known
Kern-Frenkel model~\cite{KF}, i.e. the potential between particles results from the sum of a hard-sphere potential $u_\mathrm{HS}$ and an
attractive directional interaction $u_\mathrm{patch}$. The hard-core
repulsion between two particles $i$ and $j$ is given by:

\begin{equation}
\beta u_\mathrm{HS}(r_{ij}) =
\left\{
	\begin{array}{ll}
		\infty & \mbox{if $r_{ij} < \sigma_{ij}$} \\
		\: 0 & \mbox{otherwise } 
	\end{array}
\right.,
\end{equation}

\noindent
where $\beta = 1/k_B T$, with $k_B$ Boltzmann's constant and $T$ the
temperature, $r_{ij}$ is the center-to-center distance between the
particles, and $\sigma_{ij} = (\sigma_i + \sigma_j)/2$ denotes the
contact distance between the particles, with $\sigma_i$
the hard-sphere (HS) diameter of particle $i$. The site-specific
attraction between the particles is determined by the circular patches
on the surface of each particle, which interact such that two particles
form a bond with interaction energy $\epsilon_{ij} > 0$ if  i) their
centers of mass are within a maximum interaction range $\sigma_{ij} +
\delta_{ij}$, and ii) the center-to-center vector between the particles
passes through a patch on the surface of both particles \cite{KF}. The
size of the patches is determined by an opening angle $\theta_{ij}$ (see Fig.~1).
The attractive part of the potential energy of two particles $i$ and $j$ is thus given by

\begin{eqnarray}
 u_\mathrm{patch}(\mathbf{r}_{ij}, \{\mathbf{p}_i\}, \{\mathbf{p}_j\}) = \nonumber \,\,\,\,\,\,\,\,\,\,\,\,\,\,\,\,\,\,\,\,\,\,\,\,\,\,\,\,\,\,\,\,\,\,\,\,\,\,\,\,\,\,\,\,\,\,\,\,\,\,\,\,\,\,\,\,\,\,\,\,\,\,\,\,\\
 u_\mathrm{SW}(r_{ij})  \Phi(\mathbf{r}_{ij}, \{\mathbf{p}_i\})\Phi(\mathbf{r}_{ji}, \{\mathbf{p}_j\}),
\end{eqnarray}
\noindent
where $u_\mathrm{SW}$ is a square-well attraction, given by:

\begin{eqnarray}
 u_\mathrm{SW}(r_{ij}) = 
\left\{
	\begin{array}{ll}
		-\epsilon_{ij} & \mbox{if $r_{ij} < \sigma_{ij} + \delta_{ij}$}\\
		\:\:\: 0 & \mbox{otherwise } 
	\end{array}
\right. .
\end{eqnarray}

\noindent
The function $\Phi(\mathbf{r}_{ij}, \{\mathbf{p}_i\})$ is defined as 

\begin{equation}
\Phi(\mathbf{r}_{ij},\{\mathbf{p}_i\}) =
\left\{
	\begin{array}{ll}
		1  & \mbox{if $\mathbf{\hat{r}}_{ij} \cdot \mathbf{{p}} > \cos(\theta_{ij})$ for any $\mathbf{p}$ in $\{\mathbf{p}_i\}$}\\
		0 & \mbox{otherwise } 
	\end{array}
\right.
\end{equation}

\noindent
where $\mathbf{r}_{ij} = \mathbf{r}_{j} - \mathbf{r}_{i}$,
$\{\mathbf{p}_i\}$ is a set of normalized vectors pointing from the
center of particle $i$ towards the center of each of its patches, and
$\mathbf{\hat{r}_{ij}}$ is a unit vector in the direction of $\mathbf{r}_{ij}$.

\subsection{NUMERICAL DETAILS OF THE SIMULATION}

\begin{figure}[tb]
\includegraphics[width=0.6\linewidth]{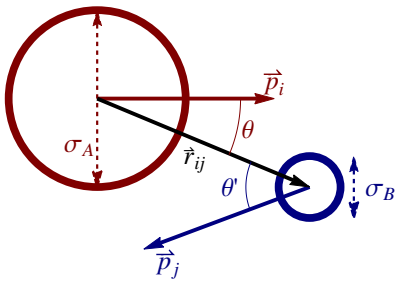}
\caption{Sketch of the geometry of the Kern-Frenkel interaction between
a particle of species $A$ and a particle of species $B$. Particles will
form an attractive bond if $r_{ij}<\sigma_{AB}+\delta_{AB}$, Eq.~(3),
and both $\cos \theta$ and $\cos \theta'$ $>\cos \theta_{AB}$, Eq.~(4). Only one of the four patches of the A particle is shown.}
\label{fig_cartoon}
\end{figure}

In the binary mixture of patchy particles studied here, the first species,
$A$, has a HS diameter $\sigma_{A}$ with four identical patches on its
surface that are arranged in a tetrahedral geometry. The second species,
$B$, has a HS diameter $\sigma_{B}=0.35\sigma_A$ and only one patch
on its surface. The attractive patch-patch interactions are modelled
by the Kern-Frenkel potential discussed above. The patch interaction
energies, Eq.~(3), for the $AB$- and $AA$-bonds are $\epsilon_{AB}$
and $\epsilon_{AA}= 0.95 \epsilon_{AB}$, respectively. Bonds between two
$B$-particles do not occur (i.e. $\epsilon_{BB} = 0$). The interaction
ranges for the patch-patch interaction are $\delta_{AA}= 0.15 \sigma_A$
and $\delta_{AB} = 0.2\sigma_A$ and the angular width of the patch
is given by $\cos \theta_{AA} = 0.92$ and $\cos \theta_{AB} = 0.99$
(see Fig. ~\ref{fig_cartoon} and Ref. \cite{inversegel}). Due to the
chosen geometry of the patches, each one can be involved only in a single
bond. In addition, the ratio $\sigma_{B}/\sigma_A$ has been chosen such
that the contribution of the $B$-particles to the total packing fraction,
close to the stoichiometric composition, was significantly smaller
than that of the $A$-particles, while maintaining the ability of the
$B$-particles to block the formation of $AA$-bonds  \cite{originale}.
The  values of the patchy interaction energies,$\epsilon_{ij}$, were
chosen to generate at low $T$ a ground state in which each $A$-particle
is bonded to four $B$-particles, completely saturating its patches.
From now on, this low $T$ preferred state will be called \textit{flower}
state. The values chosen for the  parameters that define the geometry
of the bonding volume (i.e. $\cos \theta_{ij}$ and $\delta_{ij}$, see
Appendix) reflect the requirement of having an entropy associated to the
$AA$-bonding that is larger than the one corresponding to an $AB$-bond,
thus favouring at intermediate temperature the formation of a highly
connected network of tetrahedrally coordinated $A$-particles.

We perform event-driven molecular dynamics (EDMD) simulations
\cite{rapaport,edmdanisotropic,widepatches_natphys} for systems
of $N_A=500$ and $N_B=2000$ particles (i.e. molar composition
$x_A=N_A/(N_A+N_B)=0.2$), for different total number densities
$\rho\sigma_A^3=0.5,1.0,1.5,2.0,2.5,3.0$, and $3.5$, covering a wide
range of the temperature $T\in[0.06,0.3]$  $\epsilon_{AB}/k_B$ for each
number density.  The implementation of the EDMD simulation relies on the
numerical prediction of bond formation and bond breaking events, using the
numerical techniques described in Ref. \cite{widepatches_natphys}. In
this work, the mass $m$ of each particle is taken to be the same,
setting the time unit of the simulation $\tau_0 = \sqrt{\beta
m\sigma_A^2}$. Similarly, the moments of inertia tensors of all
particles are chosen to be the same: $I_{xx} = I_{yy} = I_{zz} = m
\sigma_A^2$. During the equilibration of the simulations, the temperature
is controlled by an Andersen thermostat: periodically, randomly selected
particles are given a new translational and angular velocity, drawn from a
Maxwell-Boltzmann distribution. While measuring the dynamic properties of
the system (here the diffusion coefficient) no thermostat is used, so that
the total energy of the system is constant. To reduce equilibration times
at low $T$, we used standard $NVT$ Monte Carlo simulations ($N=N_A+N_B$)
to generate initial equilibrium configurations.

To calculate numerically the phase diagram of the binary mixture
we use the extension of the successive umbrella sampling
(SUS) method~\cite{umbrella_original}  recently developed
~\cite{lorenzo-sus-2d}. The SUS method provides an efficient way to
sample in a grand-canonical ensemble  (at fixed chemical potential,
temperature and volume) the probability that $N$ particles are present
in the simulated volume by dividing the interval $0<N<N_{max}$ (where
$N_{max}/V$ is the largest sampled density) into $N_{max}$ intervals
(in which the number of particles can fluctuate only by one unit). In
the extension to binary mixtures, the intervals $0<N_A<N_A^{max}$ and
$0<N_B<N_B^{max}$  are partitioned into overlapping two-by-two windows
($[N_A,N_A+1]$ $\times$ $[N_B,N_B+1]$) and in each of these windows a  grand
canonical Monte Carlo (GCMC) simulation is performed at fixed $V$ and $T$,
rejecting all moves which would  bring the number of particles outside
the selected range. We have selected $N_A^{max}=75$ and $N_B^{max}=320$,
corresponding to a total of 24000 distinct simulations. To improve the
method even further, the chemical potential of the two species  $\mu_A$
and $\mu_B$  is optimized in each window.  The  $\mu_A$ and $\mu_B$
values  are selected  in a preliminary run in such a way that the states
with $N_A$ and $N_A+1$ particles (as well as the states with $N_B$ and
$N_B+1$ particles) are sampled with comparable probabilities.  With this
choice, data in all windows have approximatively the same statistics. At
the end of the simulations, histogram re-weighting is performed on
the probabilities of the different windows to generate a set of data
corresponding to the same values of $\mu_A$ and $\mu_B$.  A splicing
procedure, described in detail in Ref.~\cite{lorenzo-sus-2d}, is then
used to evaluate the probability distribution function $P(N_A,N_B)$. The
resulting  $P(N_A,N_B)$ is then analysed (after re-weighting to
selected $\mu_A$ and $\mu_B$ values) to determine the density and
the relative fraction $x \equiv N_A/N_B$ of the two coexisting states
\cite{lorenzo-sus-2d}. The significant  computational investment required
is compensated by the possibility of calculating the full behavior in the
$\rho-x$ plane at the chosen $T$.  We have repeated the procedure  for
three different $T$'s (0.09$\epsilon_{AB}/k_B$, 0.12$\epsilon_{AB}/k_B$,
0.135$\epsilon_{AB}/k_B$).  Smaller $T$'s require longer equilibration
times, prohibiting us from exploring this region.

\subsection{WERTHEIM THEORY}

In this work, the analytical approach to the thermodynamics of our system
is based on Wertheim's thermodynamic perturbation theory which allows us
to obtain an analytical expression for the Helmholtz free energy, $F$,
of pure patchy particles fluids and fluid mixtures (a detailed description
can be found in Refs. \cite{wertheim1,wertheim2}). Accordingly, the total
free energy per particle, $f=F/N$, can be split into two contributions:

\begin{eqnarray}
\ f= f_{HS} + f_b \,\,, 
\end{eqnarray}
\noindent
where $f_{HS}$ is the contribution to the total free energy due to the
hard-sphere interaction whereas $f_b$ contains the contribution due to
the attractive (bonding) patchy interaction.  $f_{HS}$ can be written
as the sum of an ideal gas contribution, $f_{id}$, and an excess term,
$f_{ex}$, which accounts for the excluded volume due to the finite size
of the particles. In case of a binary mixture with molar compositions
$x_i=N_i/N$ ($i \in \lbrace A,B\rbrace$), the ideal term is given by

\begin{eqnarray}
\ \beta f_{id} = \ln(\rho\Lambda^{3}) -1 + \sum_{i=A,B} x_i\ln(x_i), 
\end{eqnarray}

\noindent where $\Lambda^3$ is the thermal volume, whose constant
contribution (that here we take as $\Lambda^3=\sigma_A^3$) is irrelevant
for the phase behavior. Concerning the excess term, $f_{ex}$, we have
used the analytical formalism developed by Mansoori and co-workers
for binary mixtures of HS \cite{Mansoori}, which takes into account the
different HS diameters of the two species (the explicit form of $f_{ex}$
is reported in the Appendix).

The bonding free energy, $f_b$, specialized to our system, is given by 

\begin{eqnarray}
\beta f_b= x_A   \left [4   \left( \ln(X_A)-\frac{X_A}{2}  \right ) + 2 \right  ] +  \nonumber \\
(1-x_A)  \left [   \left( \ln(X_B)-\frac{X_B}{2}  \right ) + \frac{1}{2} \right  ] ,  
\end{eqnarray}

\noindent
where $X_\alpha$ is the probability that a patch of type $\alpha \in
\{A,B\}$  is not bonded. These probabilities, or more conveniently
the probabilities $p_{\alpha}=1-X_{\alpha}$ that an $\alpha$-site
\textit{is bonded} can be  calculated through the law of mass
action~\cite{delasheras_soft_Wertheim} (see Appendix).

To obtain the equilibrium properties of our binary system we follow
Ref.~\cite{delasheras_soft_Wertheim} and minimized the Gibbs free
energy per particle, $g=p/\rho + f$, for a fixed pressure $p$,
$T$, and composition $x\equiv x_A$ with respect to the total number
density, $\rho$. Coexisting points are then determined by imposing
chemical equilibrium through the equality of the chemical potentials
of both species in the two coexisting phases (thermal and mechanical
equilibrium are already assured by imposing a constant $T$ and $p$). This
condition for chemical equilibrium is geometrically implemented via a
common-tangent construction of the function $g(x)$ for the two coexisting
compositions at fixed $p$ and $T$ (see Appendix).

\section{RESULTS}
\subsection{PHASE DIAGRAM: COMPOSITION VERSUS DENSITY}  
  
In this section we first discuss the composition-density profiles of the
coexistence regions at different $T$'s as predicted  by Wertheim's theory.
In general, the   coexistence region of a binary mixture defines a
three-dimensional volume in the $T-\rho-x$ space that we present here
by two-dimensional cuts in the $x-\rho$ plane for different
values of $T$.

\begin{figure}
\begin{center}
\begin{tabular}{ccc}
\hfill & \hspace{0.1cm} & \hfill \\
\includegraphics[width=0.23\textwidth]{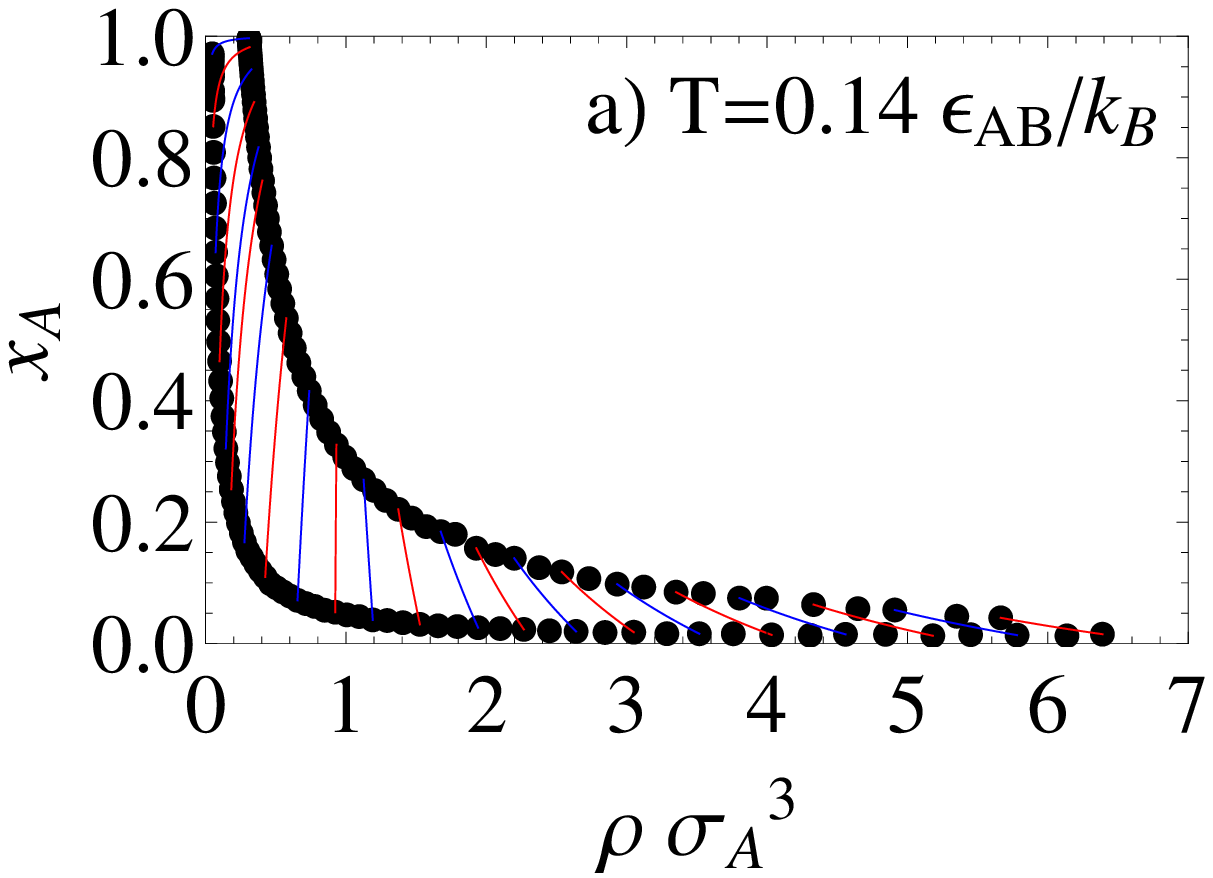} & & \includegraphics[width=0.23\textwidth]{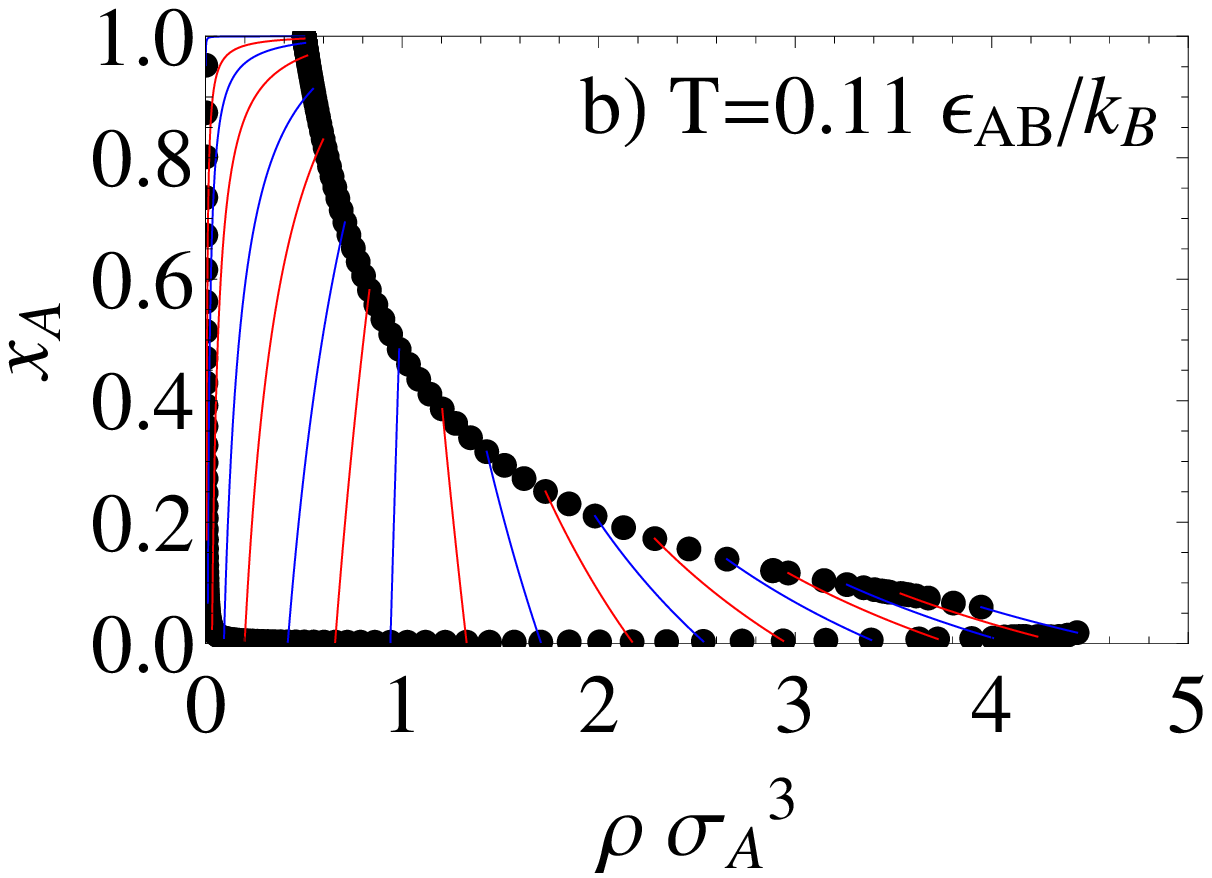} \\ 
\hfill & & \hfill\\
\includegraphics[width=0.23\textwidth]{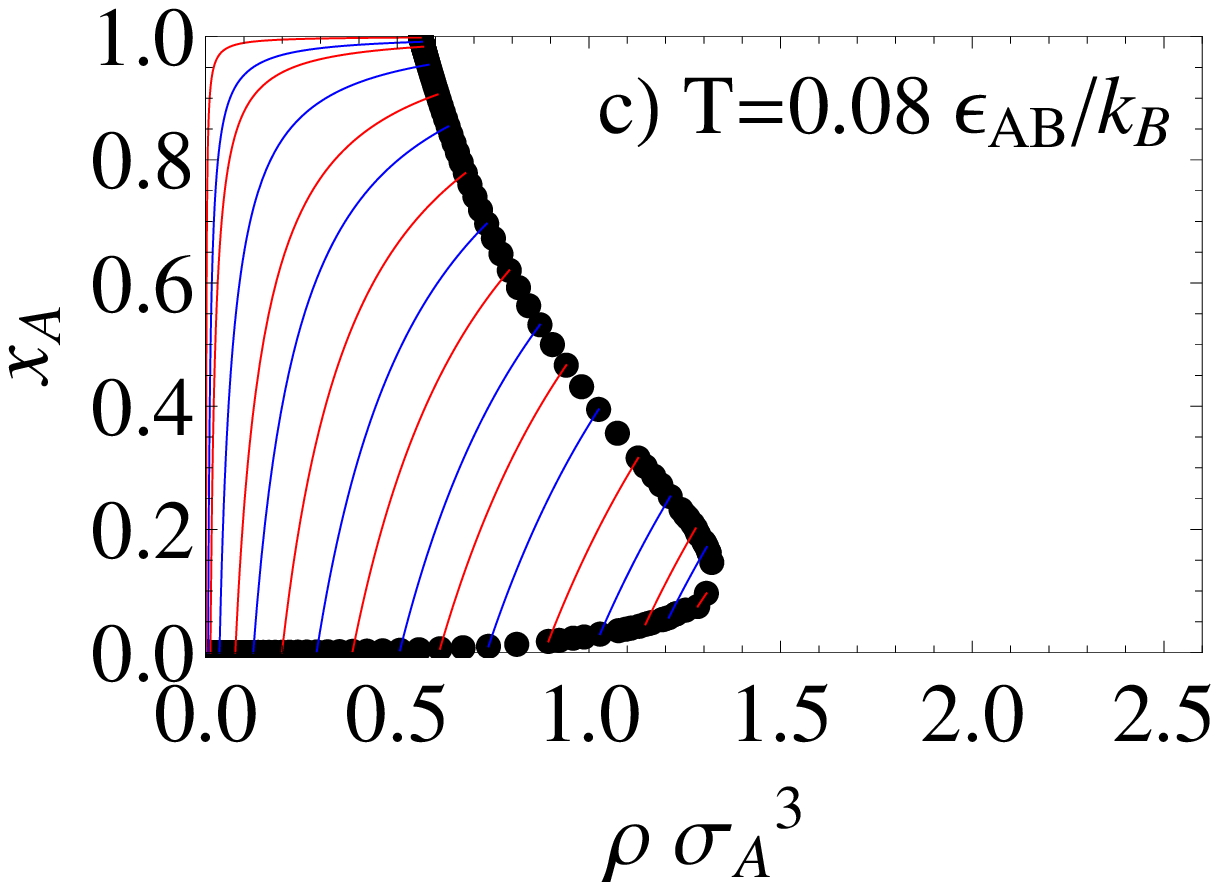} & & \includegraphics[width=0.23\textwidth]{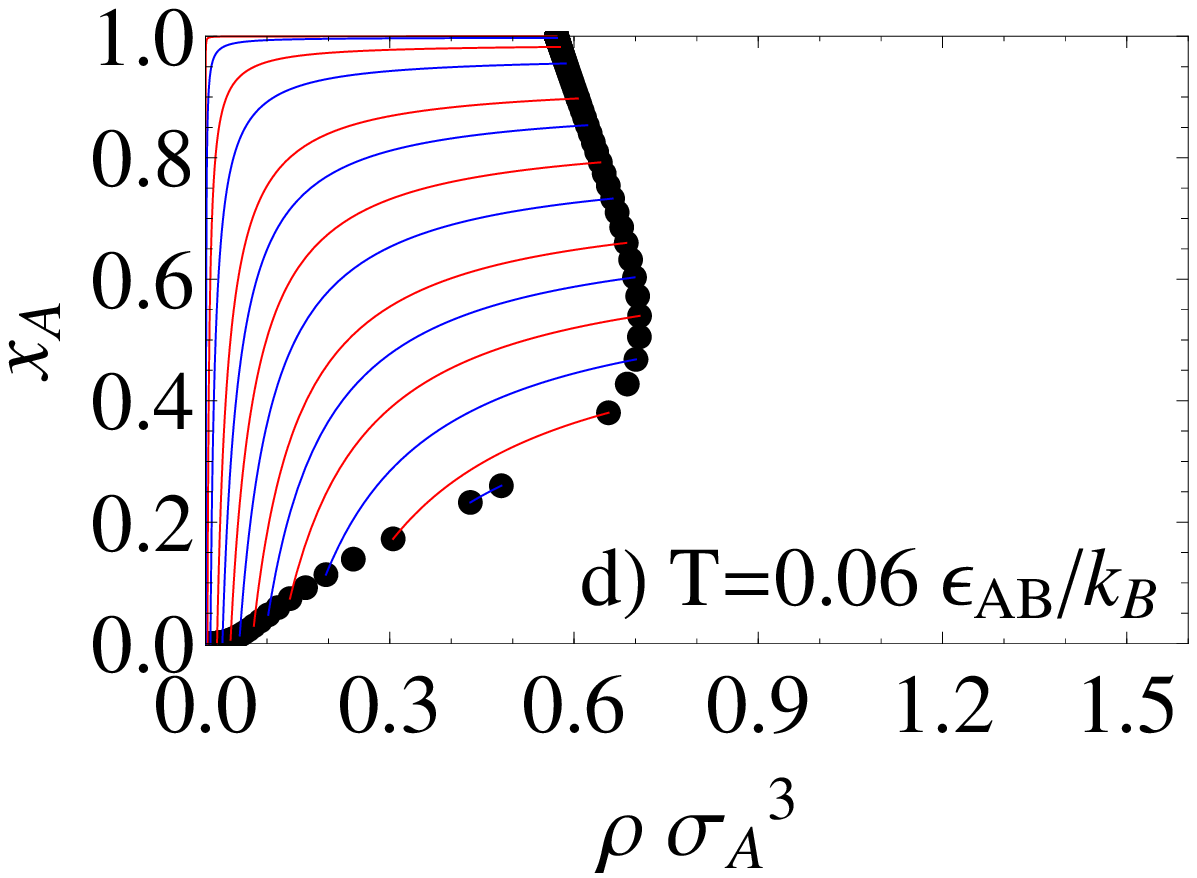} \\
\hfill & & \hfill\\
\includegraphics[width=0.23\textwidth]{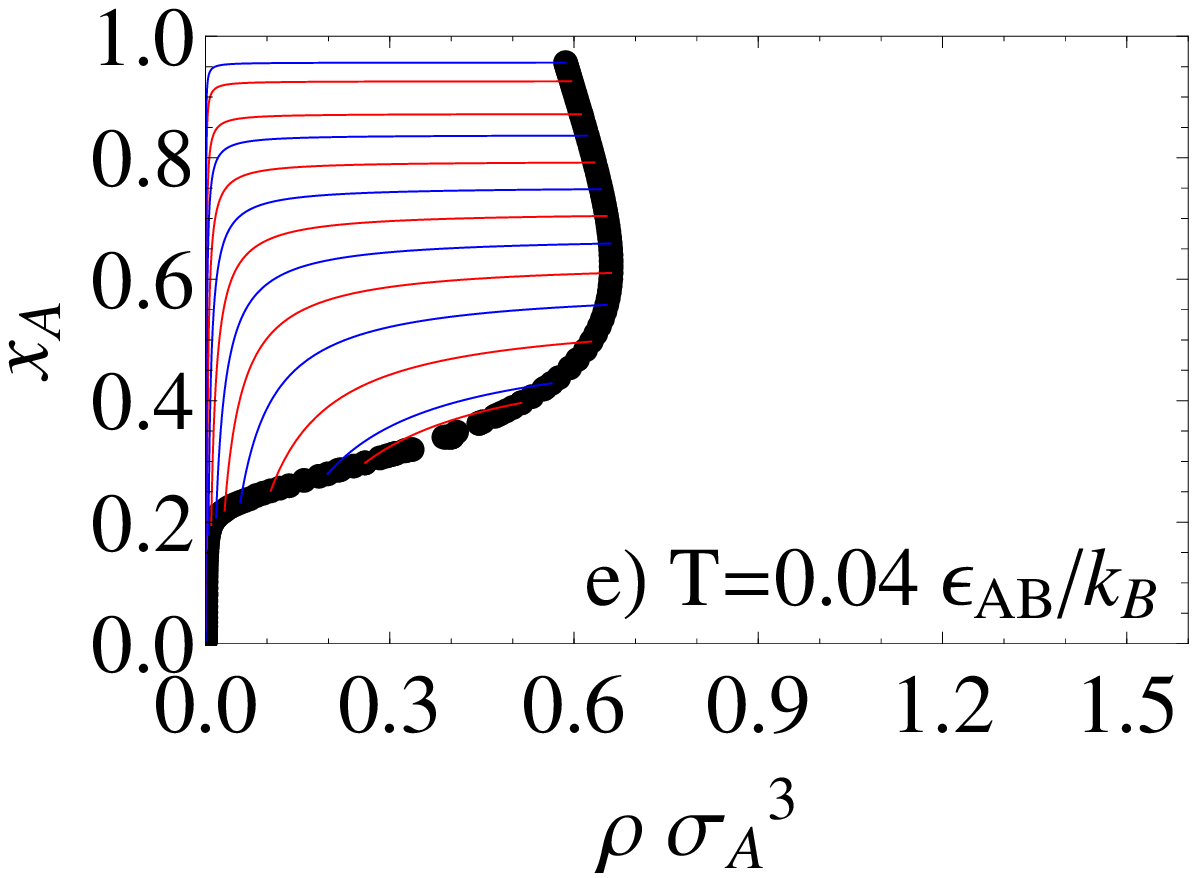} & & \includegraphics[width=0.23\textwidth]{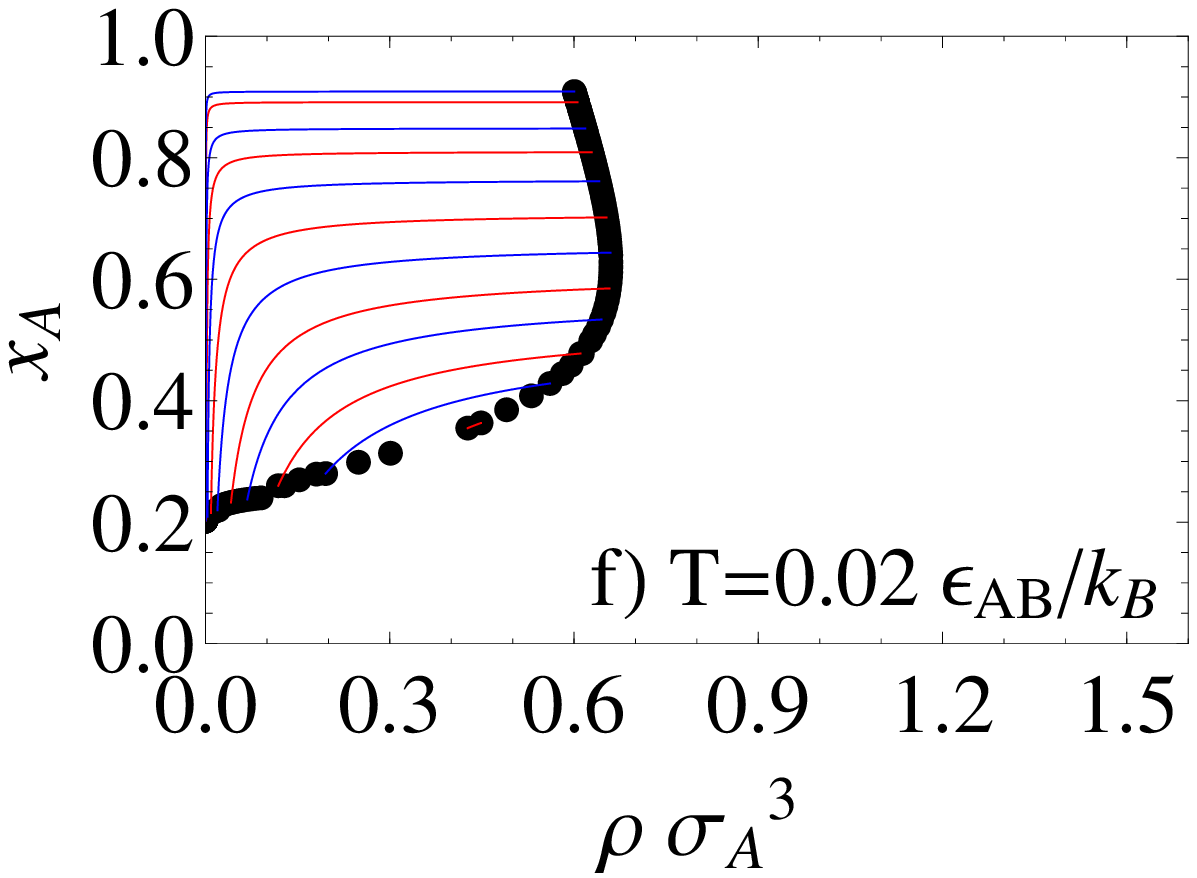} 
\end{tabular}
\caption{Two dimensional cuts of the phase diagram showing composition,
$x_A$, versus total number density of the coexistence region for
different temperatures according to Wertheim's theory. Black
dots indicate the coexisting points and their corresponding tie lines
are shown as (blue and red) solid lines.}
\label{fig_werteim}
\end{center}
\end{figure}

Figure~\ref{fig_werteim} shows the prediction of Wertheim's theory
for our system at different $T$'s. According to the theory, for
$T\leq0.15 \epsilon_{AB}/k_B$ the system phase separates into two
phases, differing in their density and/or composition, whereas for
$T>0.15 \epsilon_{AB}/k_B$ no phase separation is predicted. The
figure also shows the tie lines (see Appendix), i.e. the set of points
that separate into the same two coexisting states. Vertical tie lines
indicate those points of the phase diagram with a prevalent demixing,
that is, the two coexisting phases  only differ in their composition
but not in their total number density. Horizontal tie lines indicate
gas-liquid type separation, that is, the two coexisting phases differ in
density but not in composition.  As Fig.~\ref{fig_werteim} shows, in our
system the gas-liquid type separation only occurs close to  $x_A=1.0$
(the monodisperse case).

As we see in Fig.~\ref{fig_werteim} we expect at all $T$'s phase
separation at low densities (typically $\rho\sigma_A^3<0.6$), that would
essentially consist of two phases (see coexisting points), one rich
in $A$-particles (high $x_A$) whereas the other would mainly consist
of $B$-particles ($x_A\approx0$). This phase separation is analogous
to the one that occurs in the one component system of particles with
four-patches \cite{lorenzomolphys}, which cluster together in order to
form an $AA$-network.

At high $T$ (Figs.~\ref{fig_werteim}a) and b)), the substantial
difference between the bonding volume associated to the $AA$-bonds, ${\cal
V}_{AA}$, (see Appendix) and that corresponding to the $AB$-bonds, ${\cal
V}_{AB}$, (${\cal V}_{AA}/{\cal V}_{AB}\approx100$) plays an essential
role, making the formation of $AB$-bonds entropically unfavorable as
compared to the $AA$-bonds. Thus, the addition of $B$-particles to the
system only increases the total number density of the two coexisting
phases, without preventing phase separation. Due to excluded-volume
effects, these $B$-particles will preferentially go to the phase where
the density of $A$-particles is low, resulting in a demixing of the
two species.

The formation of clusters of $A$-particles at these temperatures
($T\sim0.11 \epsilon_{AB}/k_B$) is clearly reflected by a high
probability of forming $AA$-bonds, $p_{AA}$.  When $x_A=0.2$ this
probability is given by $p_{\rm AA} \equiv p_A-p_B$ (see Appendix
and Ref. \cite{inversegel}). Figures~\ref{fig_paa_rho}a) and b)
show $p_{AA}$ for $x_A=0.2$ at different densities as obtained by
Wertheim's theory and from our EDMD simulations respectively. Also
shown in these graphs is the probability $p_{\rm AB}=p_B$ that a patch
of type $A$ of a particle of the species $A$ is specifically bonded to a
patch of a $B$-particle. Indeed, around $T\sim0.11 \epsilon_{AB}/k_B$,
$p_{\rm AA}$ shows a maximum (Fig.~\ref{fig_paa_rho}a)), indicating
a high degree of percolation of the $A$-particles. The position of this
maximum, $T_{max}$, slightly increases upon increasing density, as can
be recognized from these figures.

\begin{figure}[tb]
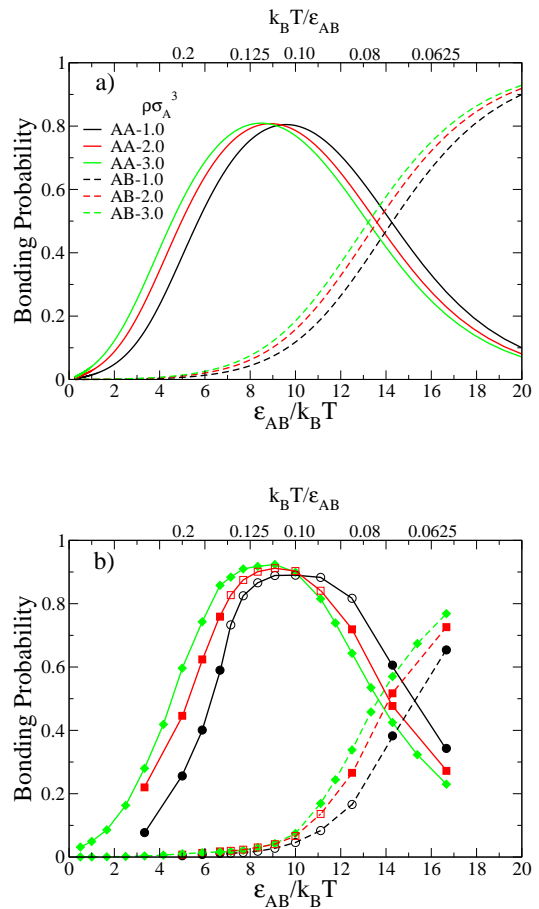

\includegraphics[width=0.8\linewidth]{fig3a_p_rho_W.eps}

\vspace{0.7cm}

\includegraphics[width=0.8\linewidth]{fig3b_p_rho_MD.eps}
\caption{a) $p_{\rm AA}$ (solid lines) and $p_{\rm AB}$ (dashed
lines) as a function of $T$ for different densities (different colors)
as obtained from Wertheim's theory for a homogeneous system with a molar
composition $x_A=0.2$. b) Same as in a) as obtained from our
EDMD simulations. Circles, squares and diamonds correspond to densities $\rho \sigma_A^3=$1.0, 2.0, and 3.0. 
Here symbols have been differentiated to account for
those states at which the system is homogeneous (full symbols) and phase
separated (empty symbols).  Note that Wertheim's theory underestimates
$p_{\rm AA}$ with respect to that obtained by EDMD in the states with a
high degree of percolation (high $p_{AA}$). This is likely due to the
presence of closed loops in the network, which are not accounted for by
the theory.}
\label{fig_paa_rho}
\end{figure}

\begin{figure}[tb]
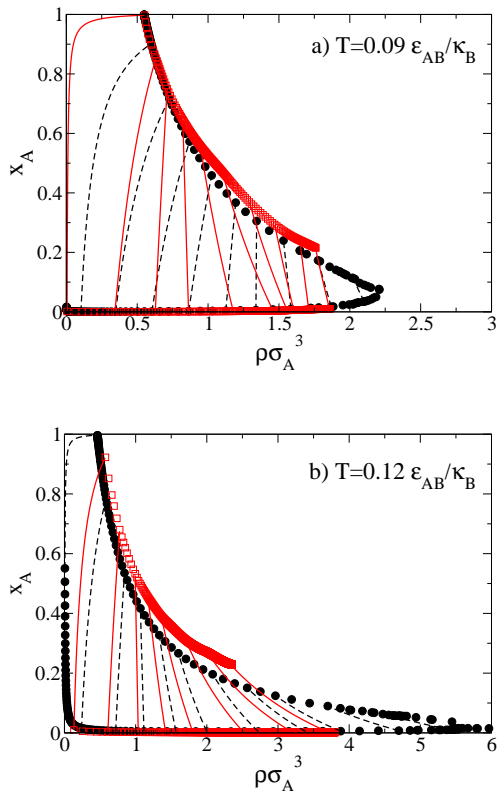

\includegraphics[width=0.75\linewidth]{fig4a_wertheim_umbrella_0.09.eps}

\vspace{0.7cm}

\includegraphics[width=0.75\linewidth]{fig4b_wertheim_umbrella_0.12.eps}
\caption{Two dimensional cuts of the phase diagram in the plane of
composition and total number density. Shown are the coexistence region
for $T=0.09 \epsilon_{AB}/k_B$ and $T=0.12 \epsilon_{AB}/k_B$ (a) and
b), respectively) according to Wertheim's theory and succesive umbrella
sampling (SUS). Black solid dots stand for the coexisting points as
obtained from Wertheim's theory whereas red empty squares represent the
coexisting points obtained by SUS. The dashed black lines are the tie
lines according to Wertheim's theory, and the red solid lines the tie
lines according to SUS.}
\label{fig_wertheim_umbrella}
\end{figure}

We see that with decreasing $T$, Figs.~\ref{fig_werteim}c)-f), the range
in density in which phase separation is observed shrinks rapidly if $x_A$
is small. This behavior is related to the fact that for $T \lesssim 0.08
\epsilon_{AB}/k_B$ the system becomes dominated by a state in which for
energetic reasons the $AB$-bonds occur very frequently.  The emergence
of this new state can be recognized from the quick increase of $p_{\rm AB}$
and the decrease of $p_{\rm AA}$, see Fig.~\ref{fig_paa_rho}. Thus for
small $x_A$ values, typically $x_A\lesssim0.2$, the high concentration
of $B$-particles limits the clustering of $A$-particles observed
at intermediate temperatures, thus driving the system back into the
homogeneous phase, Figs.~\ref{fig_werteim}d)-f). However, for high values
of $x_A$, but still low densities, the number of $B$-particles is too
small to block all of the $A$-patches. As a result, the system will still
phase separate into an $A$-rich network phase coexisting with a phase
rich in $A$-particles that are completely blocked by four $B$-particles,
i.e. that form the local structure that we referred to as \textit{flowers}
(see section II.B). In this case we have a phase separation between
two phases that differ significantly in composition and density. In
fact, this is the only phase separation that survives at very low $T$
for $\rho\sigma_A^3\lesssim0.6$ and $x_A>0.2$, where a high $x_A$ fluid
(essentially a pure $A$ fully bonded network) will coexist with a low
density fluid with $x_A \simeq 0.2$ (essentially a gas of flowers)
(Figs.~\ref{fig_werteim}e) and f)).

Next, we compare the prediction of the Wertheim theory
with the corresponding composition-density cuts through the
phase diagrams as obtained by the SUS method (see section
II.B). Figures~\ref{fig_wertheim_umbrella}a) and b) show this
comparison for two different $T$'s ($0.09 \epsilon_{AB}/k_B$ and $0.12
\epsilon_{AB}/k_B$). In both cases, the boundary of the coexistence
regions as obtained from the theory is in very good agreement with
the one obtained from the simulation, although at high densities some
differences can be seen. (Typically for $\rho\sigma_A^3>1.0$ and more
noticable at $T=0.12 \epsilon_{AB}/k_B$.) For $T=0.09 \epsilon_{AB}/k_B$,
the lowest $T$ for which we were able to equilibrate the SUS simulations,
the agreement between the boundary of the coexistence region predicted
by the theory and the one by the simulation is certainly excellent,
see Fig.~\ref{fig_wertheim_umbrella}a). However, even at this $T$ we
see that this agreement does not hold for the tie lines connecting
coexisting points at intermediate densities ($0.75 \lesssim \rho\sigma_A^3
\lesssim 1.25$).  As mentioned above, these deviations occur for those
states for which we have a well formed $AA$-network with a high degree
of percolation, i.e. geometries for which the theory is less accurate
due the presence of closed loops of particles not considered in the
theory \cite{heras:104904}.  For $T=0.12 \epsilon_{AB}/k_B$, where the
degree of percolation is slightly less pronounced, the agreement between
Wertheim and SUS tie lines is recovered for the whole set of densities
and compositions computed by the SUS method.

\subsection{THE $x_A=0.2$ PHASE DIAGRAM: TEMPERATURE VERSUS DENSITY}

In this section we discuss the 2D cut (temperature versus density) through
the phase diagram corresponding to a fixed molar composition $x_A=0.2$.
This is the so-called stoichiometric molar fraction since for this
value of $x_A$ all the $A$-patches can be bonded to $B$-patches, i.e.
the system ground state is a pure fluid of flowers. However,
we mention that this is certainly not the only interesting molar
composition, since, e.g., binary mixtures of patchy particles with
non-stoichiometric ratios \cite{frank-vitrimers} are a valuable model
for describing vitrimers, a recently invented malleable network plastic
with controlled healing properties~\cite{vitrimer-science}.

\begin{figure}[tb]
\includegraphics[width=0.75\linewidth]{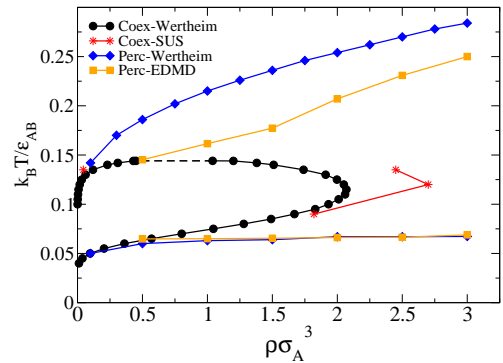}
\caption{Two dimensional cut of the phase diagram showing temperature
versus total number density for a fixed composition $x_A=0.2$. The black
dots are the coexistence points according to Wertheim's theory. The
figure also shows those coexisting points computed by succesive
umbrella sampling for $T=0.09, 0.12$, and $0.135 \epsilon_{AB}/k_B$
(red stars). Black and red solid lines connecting the points obtained
by Wertheim's theory and SUS, respectively, are guides for the eye. The
black dashed line is a guide to the eye to connect the two \textit{branches}
of the Wertheim's coexistence region. Also included are the (upper and
lower) percolation lines as obtained by Wertheim's theory (blue diamonds
connected by solid lines) and those obtained from our EDMD simulations
(orange squares connected by solid lines).}
\label{fig_T_rho}
\end{figure}

Figure~\ref{fig_T_rho} shows the coexistence region for $x_A=0.2$ in
the $T-\rho$-plane as predicted by Wertheim's theory. Also included
are the numerical results for the three temperatures using the SUS
method. According to Wertheim's theory, no phase separation can be
expected for densities $\rho\sigma_A^3\gtrsim2.0$, independently
of the temperature. Moreover, for all densities the system is
predicted to be homogeneous if $T\gtrsim0.15 \epsilon_{AB}/k_B$.
As mentioned in the previous section, the phase diagram shows a
re-entrant behavior typical of systems with competing interactions
\cite{russoj,heras:104904,frenkelnatmat}. At intermediate $T$'s
($0.03 \epsilon_{AB}/k_B \lesssim T\lesssim 0.15 \epsilon_{AB}/k_B$)
the system phase separates into a network phase mostly composed of
$AA$-bonds and another phase rich in $B$-particles. As discussed above,
this separation is essentially a demixing where the two phases have a
similar total number density but significantly different composition,
and therefore local packing fraction. (As an example, see the states
inside the coexistence region with $x_A=0.2$ in Figs.~\ref{fig_werteim}
and ~\ref{fig_wertheim_umbrella}). Indeed, one should realize that the
two branches limiting the coexistence region (black dots joined by
solid lines in Fig.~\ref{fig_T_rho}) {\it in this cut at constant
$x_A$} do not correspond to two sides of the same coexisting points
since in general the state points coexisting with those that appear in
Fig.~\ref{fig_T_rho} will have a different composition, $x_A\neq0.2$
(see Fig.~\ref{fig_werteim}). Therefore, the phase separation present
in our $x_A=0.2$ system will differ from that occurring in pure systems
of particles with valence four for which we will have a gas-liquid type
separation. However, due to the \textit{neutral} role of the $B$-particles
at high and intermediate $T$ before entering into the coexistence region,
the critical points as predicted by Whertheim's theory for our binary
mixture and for a pure $A$-system ($x_A = 1$) are similar. In our case,
and according to Wertheim's theory, the high $T$ border of the coexistence
region is placed around $T\approx 0.15 \epsilon_{AB}/k_B$ with a critical
partial density $0.1\leq\rho x_A\sigma_A^3\leq 0.2$, whereas for the pure
$A$ system we obtained $T\approxeq0.14 \epsilon_{AB}/k_B$ with a critical
density $\rho \sigma_A^3\approx 0.13$. 
We mention that these latter
values differ from those obtained by numerical simulation for pure $A$
systems with similar interaction parameters \cite{Foffi-four-valence},
where the critical point is found at $T\approx0.15 \epsilon_{AB}/k_B$
but with a significantly higher density, $\rho \sigma_A^3 \approx 0.3$. 
It is indeed well know that Wertheim theory underestimates the density of the
coexisting liquid branch~\cite{emptyliquids}.

At sufficiently low $T$, the $AB$-interactions starts to become
dominant, hence favouring the formation of inert flower structures,
and thus return the system to the homogeneous state for $T \lesssim
0.07 \epsilon_{AB}/k_B$. As already shown in the previous section,
there are discrepancies between Wertheim's theory and the SUS method at
intermediate $T$'s ($T\approx0.12 \epsilon_{AB}/k_B$) where the theory
underestimates the coexistence density of the network phase. Indeed,
at $T=0.12 \epsilon_{AB}/k_B$ and $T=0.135 \epsilon_{AB}/k_B$ the
numerical simulation shows that the range of the coexistence region
extents up to $\rho\sigma_A^3\approx2.7$. On the other hand theory and
simulation agree again for $T=0.09 \epsilon_{AB}/k_B$, i.e. when the
system becomes less percolated due to the emergence of the fluid of flowers at low $T$
($T\lesssim0.09 \epsilon_{AB}/k_B$).

Also shown in Fig.~\ref{fig_T_rho} are the percolation lines as
obtained by the Flory-Stockmayer theory \cite{florybook}, with
bonding probabilities evaluated via Wertheim's theory (always for
$x_A=0.2$). Since in our system percolation is associated to the
formation of the $AA$-network ($B$-particles merely act as blockers), the
percolation temperature for a given density will be that at which $p_{\rm
AA}=1/3$, that is, our criterion to establish the percolation temperature
of the AA-network is formally equivalent to that one that applies for one-component systems with four
valence particles \cite{flory_jacs,florybook}. We have also computed the
percolation points by directly analyzing the cluster distribution from
our EDMD simulations. In this case, we consider a system to be percolated
if the largest cluster in the system spans the simulation box. This
procedure is performed for an ensemble given by several independent
configurations. The percolation \textit{locus} is defined as the set
of state points at which at least fifty per cent of the configurations
percolate. Interestingly, the \textit{locus} of the high $T$ percolation
line obtained by Wertheim's theory overestimates the temperature at which
the percolation threshold is reached with respect to that obtained by our
EDMD simulations (see Fig.~\ref{fig_T_rho}, blue diamonds and orange squares,
respectively). This difference is likely the result of loop-formation
in the fluid (present in simulations but not in the theory). Another possible source for the deviation is the analytical
overestimation of the bonding volume ${\cal V}_{AA}$ that we performed
by using the contact value of the partial radial distribution function
between $A$-particles for the whole interaction range, $\delta_{AA}$
(see Appendix). Interestingly, the low $T$ percolation lines as obtained
from Wertheim's theory and EDMD show a very good agreement.

\begin{figure}[tb]
\begin{tabular}{cc}
a)   & b)\\
\includegraphics[width=0.45\linewidth]{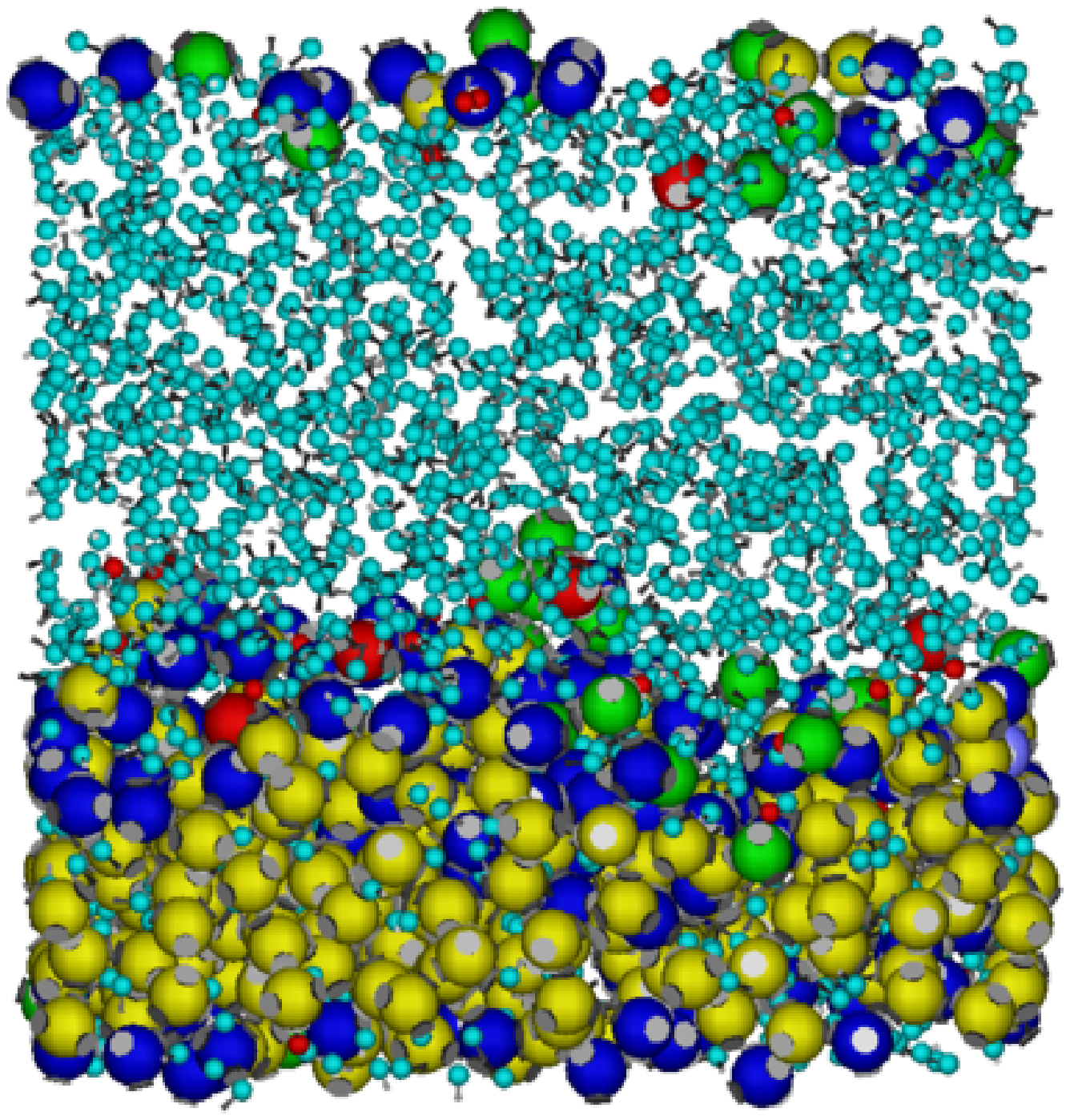} & \includegraphics[width=0.45\linewidth]{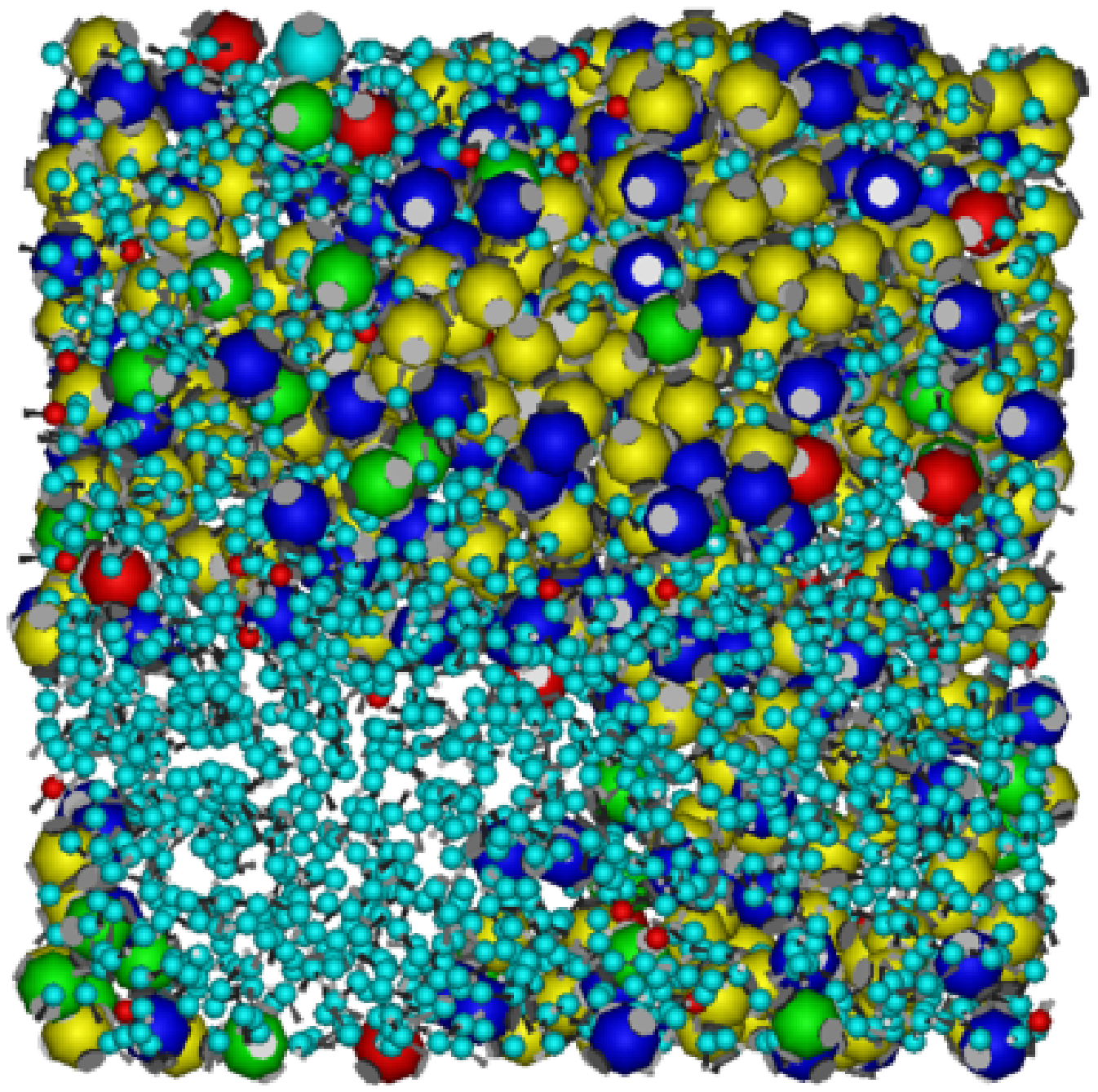} \\
c)   & d)\\  
\includegraphics[width=0.45\linewidth]{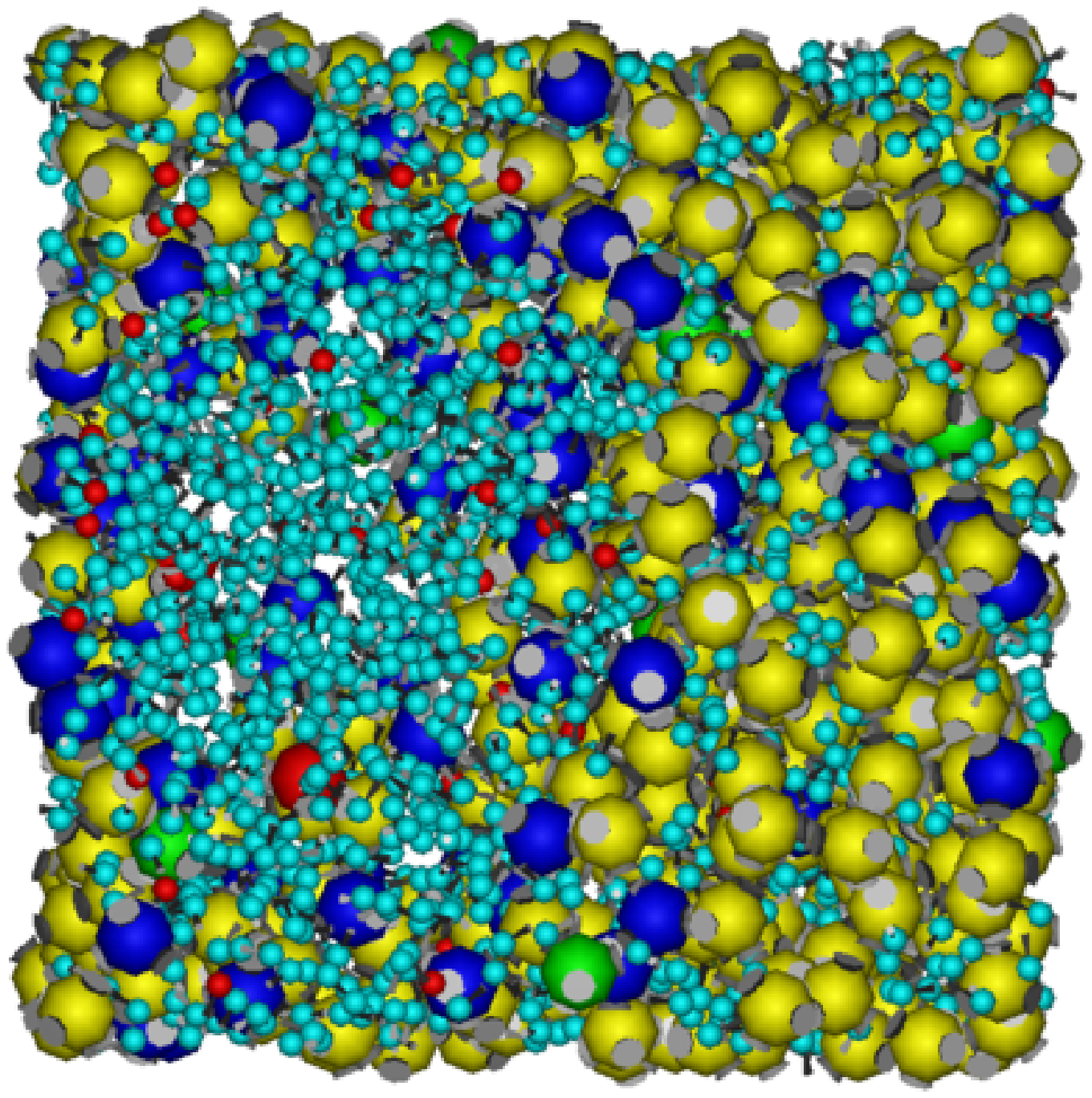} & \includegraphics[width=0.45\linewidth]{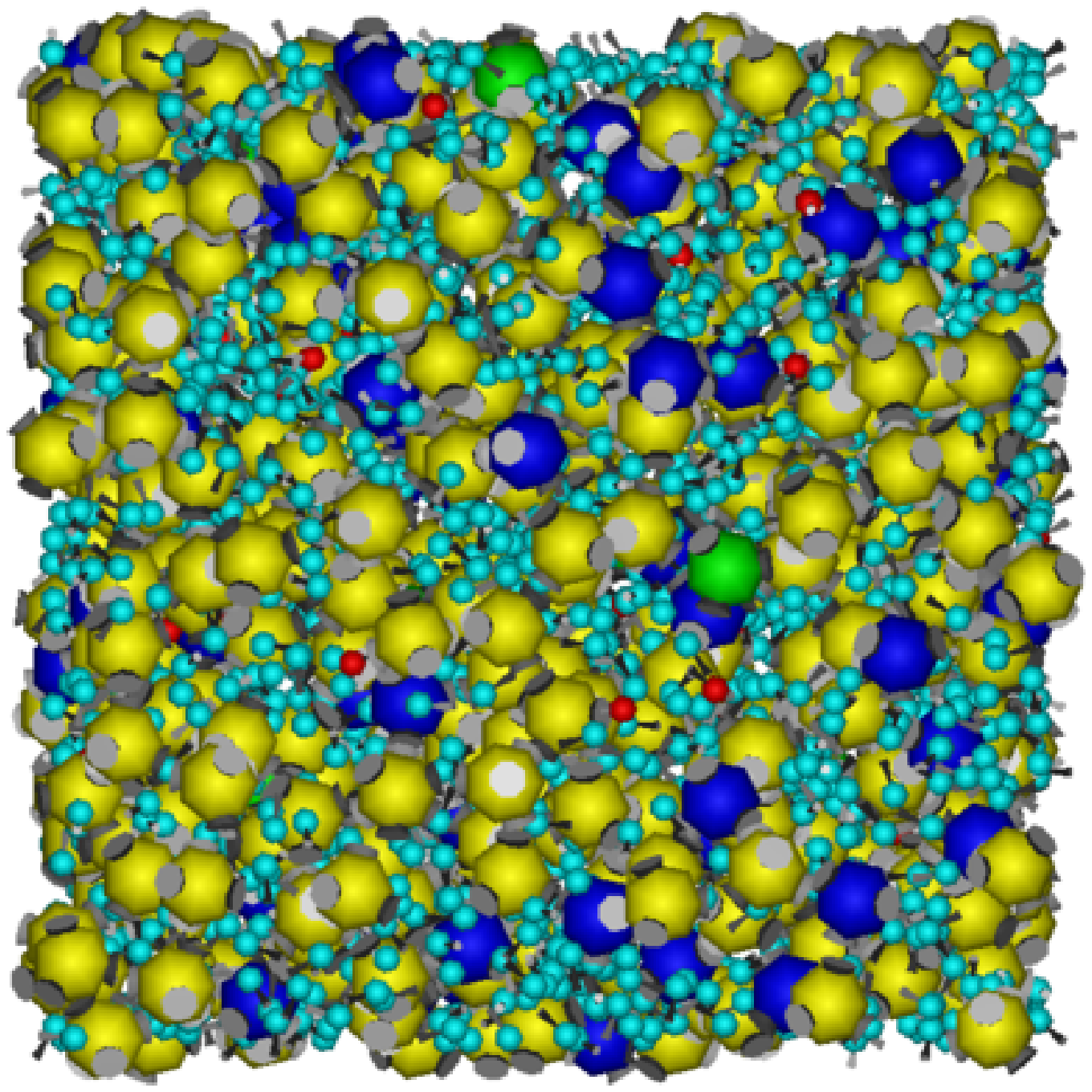}
\end{tabular}
\caption{Snapshots of four equilibrium configurations obtained from our
EDMD simulations at $T = 0.12 \epsilon_{AB} / k_B$ and $x_A=0.2$ for
four different densities: {\bf a)} $\rho \sigma_A^3 = 1.0$. {\bf b)}
$\rho \sigma_A^3 = 1.5$. {\bf c)} $\rho \sigma_A^3 = 2.0$ . {\bf d)}
$\rho \sigma_A^3 = 2.5$. Color code: Particles with four, three, two,
one, and zero bonds are displayed in yellow, blue, green, red, and light
blue respectively. Large and small sized particles correspond to the $A$-
and $B$-species respectively.}
\label{fig_snapshot_phase}
\end{figure}

Finally, we show in Fig.~\ref{fig_snapshot_phase} four snapshots
corresponding to different equilibrium configurations obtained from
our EDMD simulations for $x_A=0.2$ and $T=0.12 \epsilon_{AB}/k_B$,
the temperature at which the coexistence region extends to its maximum
density according to Wertheim's theory (see Fig.~\ref{fig_T_rho}). In
Fig.~\ref{fig_snapshot_phase} we see a clear phase separation
for densities $\rho \sigma_A^3 = 1.0$ and $\rho \sigma_A^3 = 1.5$
(Figs.~\ref{fig_snapshot_phase}a and b, respectively). As discussed
in the previous section, the separation results in two phases, one
rich in unbonded $B$-particles whereas the other mainly consists of an
$A$-particle network. For $\rho \sigma_A^3 = 2.0$, that is, the density
at which Wertheim's theory predicts the edge of the boundary of the
coexistence region, the phase separation is still obvious (see Fig.
\ref{fig_snapshot_phase}c), confirming the discrepancy between the
prediction of the simulation and Wertheim's theory. Finally, for $\rho
\sigma_A^3 = 2.5$, a density in the vicinity of the edge of the boundary
of the coexistence region as predicted by the SUS method, we only see a
small amount of phase separation, consistent with the SUS prediction. For
$\rho\sigma_A^3\gtrsim2.7$, the $x_A=0.2$ system is homogeneous for
all temperatures.

\subsection{HIGH DENSITY CRYSTAL}

\begin{figure}[tb]
\begin{tabular}{cc}
a)   & b)\\
\includegraphics[width=0.45\linewidth]{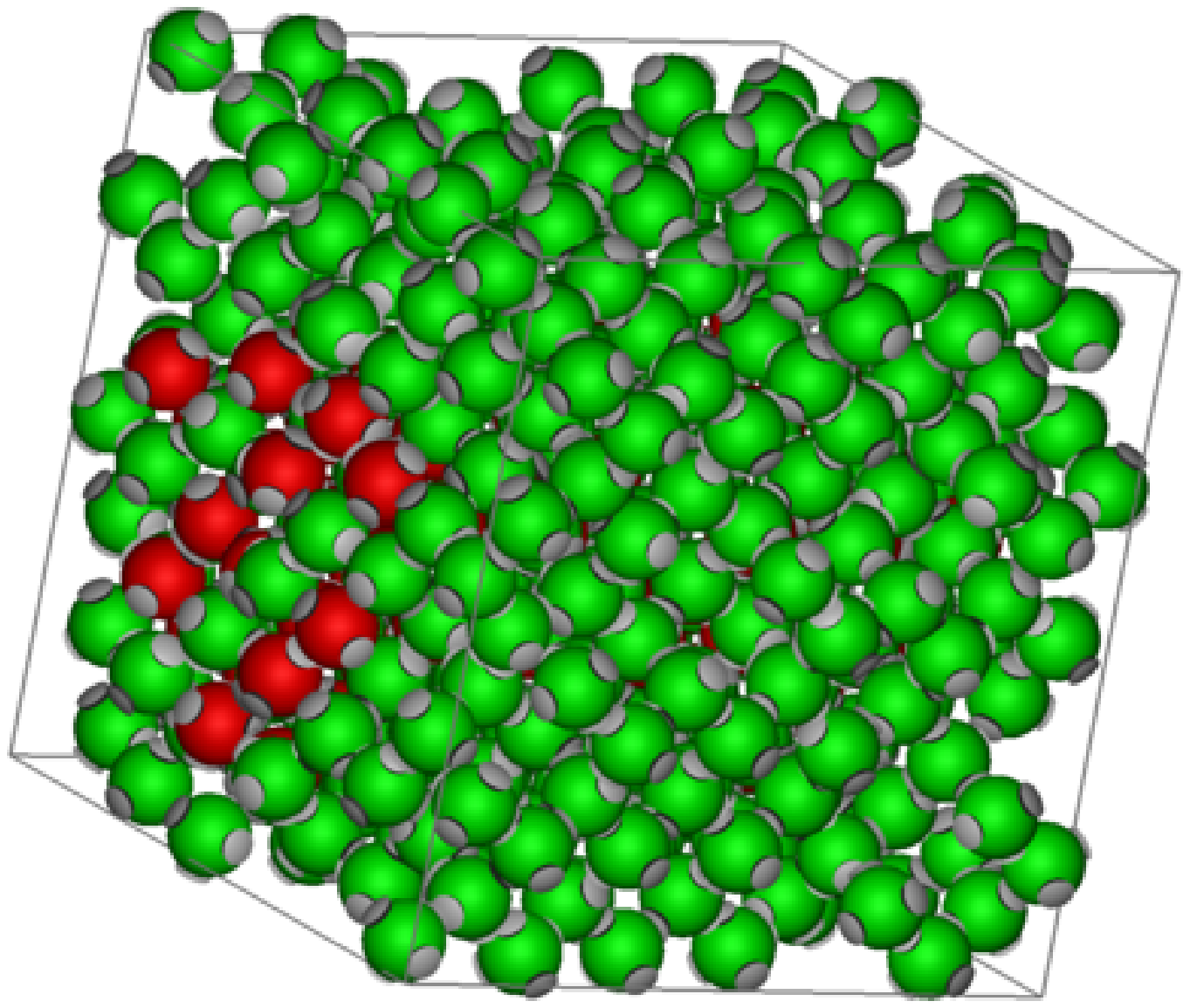} & 
\includegraphics[width=0.45\linewidth]{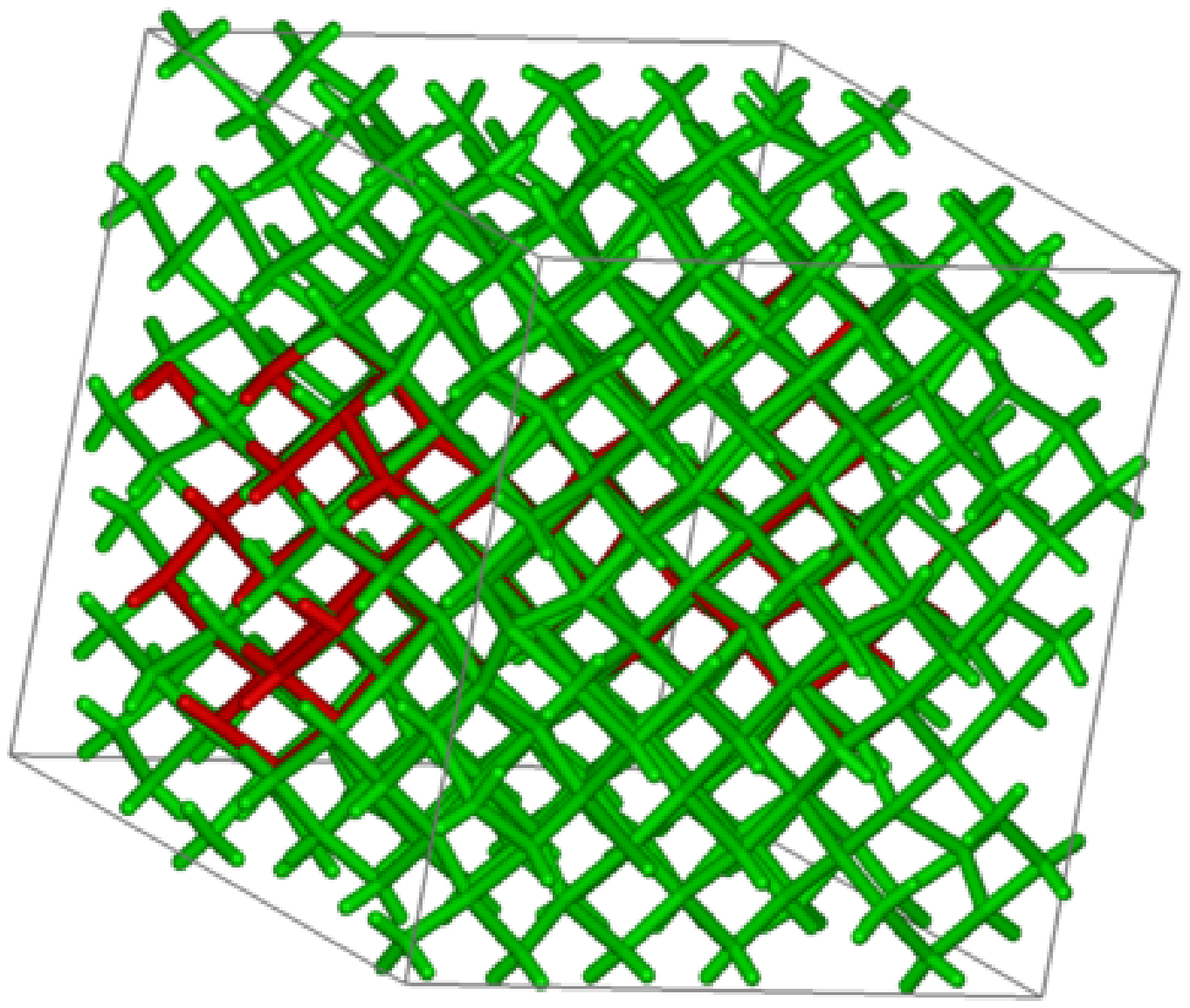}  
\end{tabular}
\caption{{\bf a)} Snapshot of a crystallized system at density $\rho
\sigma_A^3 = 3.5$, composition $x_A = 0.2$, and temperature $T =
0.13 \epsilon_{AB} / k_B$. Particles that are not part of the crystal
(including the $B$-particles) are not shown. The two colors indicate
the two separate diamond lattices, which interpenetrate to form a BCC
lattice in part of the simulation box. {\bf b)} The same configuration,
but depicting the bonds between particles instead of the particles.}
\label{fig_crystalsnapshots}
\end{figure}

As we have already mentioned, for $x_A=0.2$, intermediate temperatures
$T\approx0.10 \epsilon_{AB}/k_B$, and rather high densities
$\rho\sigma_A^3\gtrsim2.5$ our system consists of a $A$-particle bonded
network with the free smaller $B$-particles homogeneously diffusing
through the network voids (see Fig.~\ref{fig_snapshot_phase}d). An
example of this situation is the reversible gel studied in our previous
work at a constant density $\rho \sigma_A^3 = 3.0$, where a maximally
connected amorphous $AA$-network at $T\approx0.11 \epsilon_{AB}/k_B$
can be reversibly melted upon both cooling or heating the system
\cite{inversegel}. If the density is even higher (e.g. $\rho \sigma_A^3
= 3.5$, the highest density investigated) we observe a spontaneous
crystallization, at temperatures $T = 0.12 \epsilon_{AB}/k_B$, $0.13
\epsilon_{AB}/k_B$, and $0.14 \epsilon_{AB}/k_B$.

The occurring crystallization process can be followed in simulations
up to the point where most of the $A$-particles are incorporated into
a crystalline environment. The system forms a coexistence of two
crystal phases: a body-centered cubic (BCC) and diamond cubic (DC)
crystal phase, both known to be stable for monodisperse particles with
a tetrahedral patch geometry \cite{kernfrenkelphasediags}.  The BCC
phase simply consists of two interpenetrating diamond lattices in which
one of these lattices stretches through the entire simulation box,
while the BCC region is formed in the region where the second diamond
lattice is present as well. Figure \ref{fig_crystalsnapshots} shows
snapshots of the system, with the two separate lattices indicated by
different colors. Interestingly, only the cubic form of the diamond
crystal structure was formed during the nucleation of the crystal.
In all cases investigated, a cluster of the BCC crystal structure was
seen to form first. Subsequently one of the two diamond lattices inside
the crystal structure grows and eventually fills most of the simulation
box.  For temperatures between $0.12 \epsilon_{AB}/k_B \le T \le 0.17
\epsilon_{AB}/k_B$, initializing a simulation with a BCC cluster present
in the box (i.e. from a configuration taken from the nucleation process
at $T = 0.12 \epsilon_{AB}/k_B$), still results in a fully crystallized
system, suggesting that (barring finite-size effects) a crystalline
state is stable there. At lower $T$'s ($T\simeq0.11
\epsilon_{AB}/k_B$), the crystal melts, and we only observe the amorphous
gel structure.

The absence of the hexagonal diamond phase, which is known to have
a very similar free energy \cite{romano:174502}, might be
explained by the presence of the BCC crystal region, which is
not compatible with the hexagonal diamond crystal structure,
or by the depletion effect created by the small $B$-particles
\cite{private-communication-Kumar-Panagiotopoulous}. 
Indeed, the BCC part of the crystal structure always
nucleates first in our simulations. The BCC crystal region typically spans a significant part of the simulation box along at least one axis, and therefore
can affect the crystal structure in the entire volume. In much larger systems, stacking faults resulting in a mix between the cubic and 
hexagonal diamond structures might still occur far away from the BCC cluster.

It is worth noting that recent studies
have shown that tetrahedral patchy particles spontaneously crystallize in a DC or DH structure
when the patch opening, $2\theta_{AA}$, is smaller than approximately
$30^\circ$.  For the Kern-Frenkel model, this corresponds to $\cos\theta
\approx 0.96$. Larger opening angles stabilize the formation of glasses
\cite{romano:174502}, while at even larger angles the liquid becomes
thermodynamically more stable than the crystal \cite{widepatches_natphys}.
It is thus intriguing to observe spontaneous crystallization in this model
(for which $\cos\theta_{AA}=0.92)$) in the presence of a second component
($B$-particles). 

\begin{figure}[tb]
\includegraphics[width=0.7\linewidth]{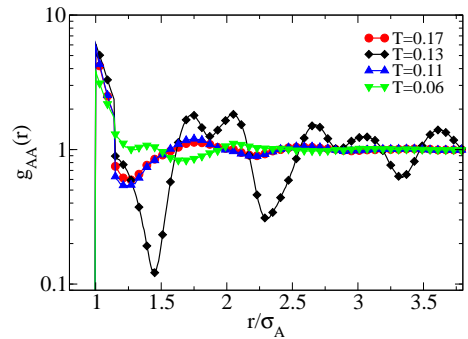}
\caption{Log-linear plot of the partial radial distribution function
$g_{AA}(r)$ for $x_A=0.2$ at density $\rho \sigma_A^3 = 3.5$. Different
temperatures (expressed in $\epsilon_{AB}/k_B$ units) are represented by
different symbols. Note that at $T = 0.13 \epsilon_{AB}/k_B$ we clearly
see the crystalline structure (black diamonds). We also mention that
all the radial distribution functions present in the figure are obtained by initializing
our simulations from non-crystallized configurations and, therefore, for $T = 0.17 \epsilon_{AB}/k_B$
we see a tetrahedral amorphous structure.}
\label{fig_gaa}
\end{figure}

The emergence of the crystalline regime and its subsequent melting is
also nicely captured through the temperature evolution of the partial
radial distribution function of the $A$-particles, $g_{AA}(r)$. Figure
\ref{fig_gaa} shows this evolution, where $g_{AA}(r)$ has been computed
at different temperatures from our EDMD simulations data. At $T = 0.17
\epsilon_{AB}/k_B$ we vaguely see the emergence of the tetrahedral
amorphous structure through an incipient second peak located around
$r=1.7\sigma_A$. At $T = 0.13 \epsilon_{AB}/k_B$ we clearly see the
signature of the crystalline phase through the well-developed maxima
and minima that extent up to large $r$ and that correspond to the
different distances that define the crystal structure. At $T = 0.11
\epsilon_{AB}/k_B$ the amorphous tetrahedral gel is again recovered,
with only  a \textit{modest} second tetrahedral peak surviving
($r\approx1.7\sigma_A$). Finally, at $T = 0.06 \epsilon_{AB}/k_B$
(the lowest temperature investigated), we see a new peak located at
around $r=1.4\sigma_A$ ($\approx \sigma_A+\sigma_B$) indicating the
emergence of a fluid composed of flowers \cite{inversegel}.

\subsection{$x_A=0.2$: TEMPERATURE AND DENSITY DEPENDENCES OF THE DIFFUSION COEFFICIENT}

\begin{figure}[tb]
\includegraphics[width=0.8\linewidth]{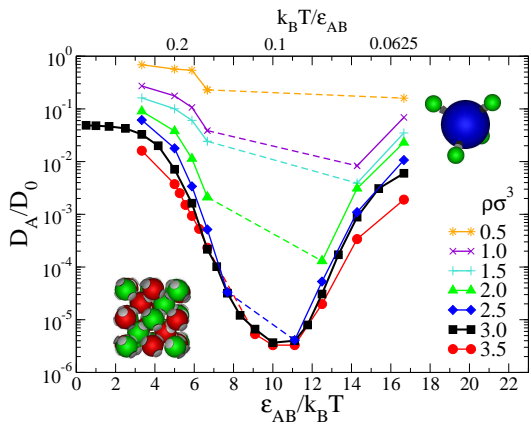}
\caption{Normalized diffusion coefficient of the $A$-particles,
$D_A/D_0$, as a function of the temperature for different densities
(symbols). The dashed lines for $\rho \sigma_A^3 \leq 2.5$ indicate
the $T$-range in which phase separation has been detected, and hence the
diffusion coefficient has not been calculated. The dashed line for $\rho
\sigma_A^3=3.5$ corresponds to the crystalline regime. Also included is
the representation of a crystal cell (bottom-left corner) that indicates
the location of the crystal regime at intermediate temperatures for
$\rho \sigma_A^3 = 3.5$. At low $T's$ the data converges toward the
diffusion constant of a fluid of flowers, the sketch of which is shown
in the upper right corner.}
\label{fig_diff}
\end{figure}

Finally we discuss the relaxation dynamics of our system and its
density and temperature dependence for the molar composition $x_A=0.2$
as obtained from our EDMD simulations. For this, we calculate the
diffusion coefficient $D_A$ of the $A$-particles from the long-time
diffusive regime of their mean square displacement (MSD) via the Einstein
relation for those densities and temperatures at which the system is
not phase separated. We recall that the center of mass of a single
species (here the $A$-species) may have a non-zero velocity, which is
compensated by the center of mass (CM) motion of the other species. When
particles form a network, like in the $AA$ case, this finite-size effect
contributes to the MSD of the single species.  To correct this effect
we subtract the CM drift of the species in question before evaluating
the MSD. In addition, to discard the trivial trend originated from the
$T$-dependence of the thermal velocity we divide $D_A$ by a reference
diffusion coefficient $D_0\equiv\sigma_A^2/\tau_0$ (see section II.B).

Figure \ref{fig_diff} shows the $T$-evolution of $D_A$ for all the
densities investigated. For  $\rho \sigma_A^3\leq2.5$, there exists a
region of intermediate temperatures where the system phase separates,
preventing the possibility of estimating the diffusion coefficient. The
$T$-range of the phase separation as seen by the omitted $D_A$ values
becomes narrower upon increasing $\rho$, as was already seen in the
phase diagram of Fig.~\ref{fig_T_rho}. Similarly, at large densities
($\rho \sigma_A^3=3.5$), the system crystallizes at intermediate $T$,
again preventing the possibility of covering the complete $T$-range of
the diffusion coefficient. Fortunately, at intermediate densities ($\rho
\sigma_A^3\approx3.0$), it is possible to follow the entire dynamical
process and its non-monotonic $T$ evolution. Figure \ref{fig_diff} clearly
shows that, on cooling, the diffusion coefficient of the $A$-particles
first decreases by four orders of magnitude and then increases going back
to typical fluid-like values.  Reference \cite{inversegel} discusses
in detail this process, associated to the progressive formation of the
$AA$-network which is then replaced by the gel-fractioning induced by the
progressive formation of $AB$-bonds. This behavior can be rationalized
by a competitive interaction mechanism that in the present case leads
to an entropically favorable $AA$-bonding at intermediate temperatures
and to an energetically favorable $AB$-bonding at low temperatures.

\begin{figure}[tb]
\includegraphics[width=0.75\linewidth]{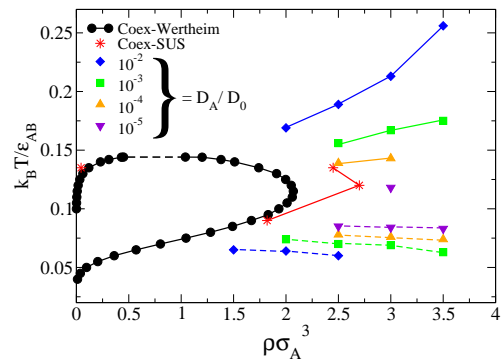}
\caption{Iso-diffusivity points for different values of $D_A/D_0$. Points
with the same value of $D_A/D_0$ are represented with a common symbol and
connected by solid (dashed) lines for the high (low) $T$ regime. Also
included is the coexistence region as obtained by Wertheim's theory
(black solid dots) and those coexisting points computed by umbrella
sampling (red stars) which have been connected by red solid lines to
gain intuition on the \textit{true} extension of the coexistence region
(see Fig.~\ref{fig_T_rho}).}
\label{fig_iso_Diffusivity}
\end{figure}

Finally, Fig.~\ref{fig_iso_Diffusivity} shows iso-diffusivity
points \cite{Foffi-iso,zacca-iso} corresponding to different $D_A$
values. At intermediate $T$ (typically $T>0.15 \epsilon_{AB}/k_B$)
we find the expected behavior of a fluid-like system: The curves have
a positive slope, i.e. states with the same value of the diffusion
coefficient have a higher temperature if density is increased (points
connected by solid lines in Fig.~\ref{fig_iso_Diffusivity}). However,
at low $T$  the opposite behavior appears, that is, states sharing a
common value of the diffusion coefficient have a slightly {\it lower}
$T$ value as density increases (points connected by dashed lines in
Fig.~\ref{fig_iso_Diffusivity}). This {\it a priori} odd behavior can
be understood by considering the density dependence of $p_{\rm AB}$
(or $p_{\rm AA}$) at low $T$ (Figs.~\ref{fig_paa_rho}a and b). From these
figures we recognize that at a fixed (low) $T$, $p_{\rm AB}$ increases as
density increases, indicating that for a fixed value of the temperature
the system will be more fragmented at higher densities. This increasing
degree of fragmentation causes the system to diffuse faster as $\rho$
increases. 

\section{CONCLUSIONS} 

In this work we have presented an extensive study of the phase diagram of
a binary mixture of $A$ and $B$ patchy particles which was specially
conceived to form a reversible gel upon both cooling and heating
\cite{inversegel}. The design of the interaction geometry and
the choice of the interaction energies results in a competing mechanism
between an entropically favorable network at intermediate temperatures
which is formed by $A$-particles (the gel) and an energetically favorable
fluid composed of $A$-particles completely bonded to $B$-particles
at low temperatures. To explore the phase diagram of the system (and
therefore its coexistence region), we have covered both analytically and
by means of computer simulations a wide range of temperature, density,
and composition.

In the analytical approach to the thermodynamics of the system we have
used a hard-sphere expression for the free energy (in which we have taken
into account the different hard-sphere diameters of the two species) and
added a bonding free energy contribution based on Wertheim's first-order
perturbation theory \cite{Mansoori,wertheim1,wertheim2}. This approach
provided us with an analytical expression for the Helmholtz free
energy and therefore with the full thermodynamics of our system. In
addition, we have performed two kinds of computer simulations for
the same model. On one hand, we have used grand canonical Monte Carlo
simulations to numerically calculate the phase diagram of our model at
different temperatures by using an extension of the successive umbrella
sampling (SUS) technique for binary mixtures \cite{lorenzo-sus-2d}. On the
other hand, to study also the dynamics of our system, we have performed
event-driven molecular dynamics simulations for a fixed molar composition
$x_A=0.2$, which represents the stoichiometric molar fraction in our
system for which all the $A$-patches can be bonded to $B$-particles.

Our calculations have allowed us to analyse the 2D cuts, composition
versus density, of the phase diagram at different fixed temperatures
as obtained from Wertheim's theory and SUS. States at which the system
is phase separated usually consist of two phases that are demixed: A
network phase rich in $A$-particles and a fluid rich in $B$-particles. For
small $x_A$ values ($x_A<0.2$), this phase separation extends to higher
densities upon increasing temperature. At low $T$, phase separation
only takes place at $x_A>0.2$ and low densities. The coexistence
regions as predicted by Wertheim's theory and SUS showed in general a
good quantitative agreement, with some deviations for those states at
which we have a high degree of connectivity of the $A$-particles. In
particular, while Wertheim's theory predicts no phase coexistence for
$\rho \sigma_A^3 > 2.0$ at any $T$, the SUS method extends the coexistence
region to $\rho \sigma_A^3 \approx 2.7$.

Using EDMD simulations, we have specifically studied the composition $x_A=0.2$,
confirming the extent of the coexistence region predicted by SUS until
$\rho \sigma_A^3 \approx 2.7$. In addition, we have also established the
temperature above which, independent of density, no phase separation is expected,
finding it to be around $T \approx 0.15 \epsilon_{AB}/k_B$. Since in our
binary system phase separation is mainly due to demixing, in general the boundary
of its coexistence region differs from that corresponding to
a pure $A$ system for which we can only have a gas-liquid type separation
\cite{Foffi-four-valence,emptyliquids,bianchi_jcp}. Also for this
composition, we have shown that the percolation lines as obtained from
Wertheim's theory and from our EDMD simulations are in good qualitative
agreement, showing a large density range where the systems only percolates
in the intermediate temperature regime. In other words, upon cooling,
the system first forms a percolating network, that then fragments again
upon further cooling.

For $x_A=0.2$ we have also investigated the spontaneous crystallization of
our system at high densities ($\rho \sigma_A^3 = 3.5$) and intermediate
temperatures ($T \approx 0.13 \epsilon_{AB}/k_B$) as well as its
subsequent melting recovering the gel state upon decreasing $T$. In
this crystalline regime, the system spontaneously forms a coexistence
of two crystal phases: a body-centered cubic (BCC) and a diamond cubic
(DC) crystal phase. Interestingly, it appears that stacking faults
associated to the hexagonal and cubic forms of diamond are not observed
\cite{romano:174502}.

Finally, we have studied the relaxation dynamics of our system
through the diffusion coefficient of the $A$-particles for $x_A=0.2$
at those temperatures and densities for which no phase separation or
crystallization is present. We have shown that for densities around
$\rho \sigma_A^3 \approx 3.0$ we can follow the complete non-monotonic
temperature evolution of the diffusion coefficient.

In summary, we have presented an extensive study of the thermodynamics
and dynamics of a reversible gel of patchy particles by using different
numerical simulations complemented with an analytical approach. With
these techniques we have explored the $T-\rho-x$ space that defines
the three-dimensional phase diagram of our system, locating its
coexistence region and describing its different phases. In particular,
we have presented a promising result for future experimental realizations
\cite{dnatetramersJACS,dnabellini,dnapatchy1} by demonstrating that there
exists a \textit{broad} density range where our system exhibits the
phenomenology of a thermo-reversible gel that can be fluidized by both
cooling and heating in the absence of phase separation.

\section{ACKNOWLEDGEMENTS} 
We acknowledge support from COMPLOIDS, ERC-PATCHYCOLLOIDS-226207 and
MIUR-PRIN. W. Kob acknowledges support from the Institut Universitaire de
France. We thank L. Rovigatti for providing us with the code to perform
and analyze the SUS simulations.

\section{APPENDIX: FREE ENERGY OF A BINARY MIXTURE OF HARD-SPHERES WITH
PATCHY INTERACTIONS.} As mentioned in section II.C, the Helmholtz free
energy of a binary mixture of HS, $f_{HS}$, can be separated into an ideal
gas contribution, $f_{id}$, and an excess term, $f_{ex}$ (see Eqs.
(5) and (6)). In our study we have considered the different particle HS
diameters following the approach of Mansoori and co-workers~\cite{Mansoori}
in which $f_{ex}$ is written as:

\begin{eqnarray}
\ \beta f_{ex}= -\frac{3}{2}(1-y_1+y_2+y_3)+(3y_2+2y_3)(1-\xi)^{-1}+ \nonumber 
\\ +\frac{3}{2}(1-y_1-y_2-y_3/3)(1-\xi)^{-2}+ \nonumber \,\,\,\,\,\,\,\,\,\,\,\,\,\,\,\,\,\,\,\,\,\,\,\,\,\,\,\,\
\\ +(y_3-1)\ln(1-\xi) \,\,\,\,\,\,\,\,\,\,\,\,\,\,\,\,\,\,\,\,\,\,\,\,\,\,\,\,\,\,\,\,\,\,\,\,\,\,\,\,\,\,\,\,\,\,\,\,\,\,\,\,\,\,\,\,\,\,\,\,\,\,\,\,\,\,\,\,\,\,\,\,\
\end{eqnarray}

where:

\begin{eqnarray}
\ \xi=\sum_{i=A,B} \xi_i \,\,\,\,\  ; \,\,\,\  \xi_i=\frac{\pi}{6}\rho x_i\sigma_i^{3} \,\,\,\,\,\,\,\,\,\,\,\,\,\,\,\,\,\,\,\,\,\,\,\,\,\,\,\,\
\end{eqnarray}

\begin{eqnarray}
\ y_1=\Omega_{AB}(\sigma_A+\sigma_B)(\sigma_A\sigma_B)^{-1/2} \,\,\,\,\,\,\,\,\,\,\,\,\,\,\,\,\,\,\,\,\,\
\end{eqnarray}

\begin{eqnarray}
\ y_2=\Omega_{AB}\frac{(\sigma_A\sigma_B)^{1/2}}{\xi}\Bigl(\frac{\xi_A}{\sigma_A}+\frac{\xi_B}{\sigma_B}\Bigr) \,\,\,\,\,\,\,\,\,\,\,\,\,\,\,\,\,\,\,\
\end{eqnarray}

\begin{eqnarray}
\ y_3=\Bigl[\Bigl(\frac{\xi_A}{\xi}\Bigr)^{2/3}x_A^{1/3}+\Bigl(\frac{\xi_B}{\xi}\Bigr)^{2/3}x_B^{1/3}\Bigl]^3 \,\,\,\,\,\,\,\,\,\,\
\end{eqnarray}

and 

\begin{eqnarray}
\ \Omega_{AB}=\Bigl[\frac{(\xi_A\ \xi_B)^{1/2}}{\xi}\Bigr]\Bigl[\frac{(\sigma_A\ - \sigma_B)^2}{\sigma_A\ \sigma_B}\Bigr](x_A x_B)^{1/2} \,\, .
\end{eqnarray}

Here we have followed the same notation used in the main text for the
total number density, $\rho$, the particle diameter, $\sigma_i$, and
the molar composition, $x_i=N_i/N$, where $i \in \{A,B\}$.

Apart from the HS contribution to the total free energy, we have
an additional contribution due to the bonding free energy, $f_b$,
discussed in section II.C, which is a function of the probabilities
$p_{\alpha}$ for $\alpha\in \{A,B\}$ (see section II.C, Eqn.
(7)). These probabilities are obtained through the law of mass action
\cite{delasheras_soft_Wertheim} that in our case takes the form of two
coupled non-linear equations:

\begin{eqnarray}
\ p_{A}=& 1- \Bigl[1+\rho \sigma_A^3[4 x_{A}  (1-p_{A})\Delta_{AA}+ \Bigr.  \nonumber \\
   &  \Bigl. + (1-x_{A}) (1-p_{B})\Delta_{AB}]\Bigr]^{-1}    
\end{eqnarray}

\begin{eqnarray}
\ p_{B}=1-\Bigl[1+\rho\sigma_A^3[4x_{A}(1-p_{A})\Delta_{AB}]\Bigr]^{-1},
\end{eqnarray}

\noindent
where all the interaction parameters needed for describing bonding
between $AA$- and $AB$-patches enter in $\Delta_{AA}$ and $\Delta_{AB}$
\cite{delasheras_soft_Wertheim}:

\begin{eqnarray}
\ \Delta_{AA}=g_{AA}(\sigma_A)[\exp{(\epsilon_{AA}/k_B T})-1]{\cal V}
_{AA}/\sigma_A^3     
\end{eqnarray}
\begin{eqnarray}
\ \Delta_{AB}=g_{AB}(\sigma_{AB})[\exp{(\epsilon_{AB}/k_B T)}-1]{\cal V}_{AB}/\sigma_A^3
\end{eqnarray}

\vspace{0.1cm}
In our work we have approximated $\Delta_{AA}$ and $\Delta_{AB}$ by
using the contact values of the partial radial distribution functions,
$g_{AA}(\sigma_A)$, $g_{BB}(\sigma_B)$, and $g_{AB}(\sigma_{AB})$
(where $\sigma_{AB}=(\sigma_A+\sigma_B)/2)$, for a binary mixture of hard
spheres as obtained from the Percus-Yevick Equation \cite{lebowitz_PY_BM}:

\begin{eqnarray}
\ g_{\alpha\alpha}(\sigma_\alpha)=\lbrace (1-\xi)+\frac{3}{2}\sigma_\alpha \chi\rbrace (1-\xi)^{-2} \, , \, \alpha\in\{A,B\} \,\,\,\,\,\,\,\,\,\ 
\end{eqnarray}
\begin{eqnarray}
\ g_{AB}(\sigma_{AB})=\Bigl[\sigma_B g_{AA}(\sigma_A) +\sigma_A g_{BB}(\sigma_B)\Bigr]/2\sigma_{AB} ,\,\,\,\,\,\
\end{eqnarray}

\noindent
where:
\vspace{0.1cm}
\begin{eqnarray}
\ \chi=\frac{\pi}{6}(\rho x_A\sigma_A^{2}+\rho x_B\sigma_B^{2}) \quad .
\end{eqnarray}
\vspace{0.1cm}

The bonding volumes ${\cal V}_{AA}$ and ${\cal V}_{AB}$ present in Eqs. (16) and (17) are given by:

\begin{eqnarray}
\ {\cal V}_{AA}=\frac{4\pi}{3}\Bigl( \frac{1-\cos{\theta_{AA}}}{2}\Bigr)^2 \Bigl[(\sigma_A+\delta_{AA})^3-\sigma_A^3\Bigr] \,\,\,\,\,\,    
\end{eqnarray}
\begin{eqnarray}
\ {\cal V}_{AB}=\frac{4\pi}{3}\Bigl(\frac{1-\cos{\theta_{AB}}}{2}\Bigr)^2\times \nonumber \,\,\,\,\,\,\,\,\,\,\,\,\,\,\,\,\,\,\,\,\,\,\,\,\,\,\,\,\,\,\,\,\,\,\,\,\,\,\,\,\,\,\,\,\,\,\,\,\, \\ 
\times \Bigl[(\sigma_{AB}+\delta_{AB})^3-\sigma_{AB}^3\Bigr],\,\,\,\
\end{eqnarray}

\noindent
where $\delta_\gamma$ and $\theta_\gamma$ ($\gamma\in\{AA,AB\}$)
are respectively the interaction ranges and the angular patch widths
presented in sections II.A and II.B. With our choices for the interaction
parameters we have ${\cal V}_{AA}=3.49\cdot 10^{-3} \sigma_A^{3} \gg
{\cal V}_{AB}=3.79\cdot 10^{-5} \sigma_A^{3}$.  Once Eqs. (14) and (15)
are solved (by using Eqs. (16)-(22)), the probabilities $p_{\rm AA}$
and $p_{\rm AB}$ that an $A$-site is specifically bonded to another $A$-
or to a $B$-site are obtained by the relations: $p_{\rm AA}=p_A-p_B$
and $p_{\rm AB}=p_B$.

We finally present the procedure for constructing the phase diagram of
our binary system, i.e locating its coexisting points, through
the minimization of the Gibbs free energy once we have the analytical
expression for the total Helmholtz free energy, $f$. First we obtain,
for a fixed composition $x^f (\equiv x_A^f)$, temperature $T^f$, and
for a given (target) pressure $p^*$, those densities that satisfy the
equation of state:

\begin{eqnarray}
\ p(x^f, T^f; \rho) = \rho^2 \left(\frac{\partial f}{\partial \rho} \right)_{x^f, T^f} = p^* 
\end{eqnarray}

In principle, we can have different densities which are solution of
Eqn.~(23) (all of them having a common pressure, composition, and
temperature). From these densities we take that one which minimizes the
Gibbs free energy per particle of our system, $g(x^f,T^f,p^*;\rho)=p^*/\rho+f$, i.e. we
take that density for which our system is most stable. This process is
then repeated for a new composition by keeping $p^*$ and $T^f$ constant,
thus covering the whole range of compositions, $x\in[0,1]$. As a result
we obtain the stable values of the Gibbs free energy $g(x)_{T^f,p^*}$
for any composition $x\equiv x_A$ (and therefore for any partial density,
$\rho_A=x_A\rho$) for a fixed pressure and temperature.

\begin{figure}[tb]
\includegraphics[width=0.8\linewidth]{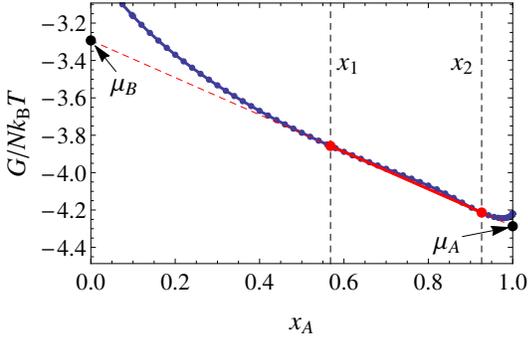}
\caption{Gibbs free energy per particle for a fixed temperature and
pressure as a function of the composition $x\equiv x_A$ (blue dots
with a solid blue line). Compositions $x_1$ and $x_2$ of the two
coexisting phases that share a common tangent have been highlighted by
vertical dashed lines. The common tangent that crosses $(x_1,g(x_1))$
and $(x_2,g(x_2))$ (red dots) has been represented by a red line. Also
marked in the figure are the values of the chemical potentials of the
two species $\mu_A$ and $\mu_B$ for the two coexisting points.}
\label{fig_commontangent}
\end{figure}

Figure \ref{fig_commontangent} shows an example of $g(x \equiv
x_A)_{T^f,p^*}$ at $T^f= 0.14 \epsilon_{AB}/k_B$ in which we have two
points $x_1$ and $x_2$ that share a common tangent, that is:

\begin{eqnarray}
\  \frac{\partial g(x)}{\partial x}\Big\vert _{x=x_1}=\frac{\partial g(x)}{\partial x}\Big\vert _{x=x_2} 
\end{eqnarray}
\\
\begin{eqnarray}
\ \frac{\partial g(x)}{\partial x}\Big\vert _{x=x_1}=\frac{g(x_1)-g(x_2)}{x_1-x_2} \,\,\,\,\,\,\,\,\,\,\,\,\
\end{eqnarray}
\\
\noindent
In Eqs. (24) and (25) we recognize the conditions of chemical
equilibrium of two phases with compositions $x_1$ and $x_2$, corresponding
to the total number densities $\rho_1$ and $\rho_2$ which are the
solutions of Eqn.~(23) for $x^f=x_1$ and $x^f=x_2$. In other words,
Eqs.~(24) and (25) are equivalent to the equalities of the chemical
potentials $\mu_A$ and $\mu_B$ of the two species $A$ and $B$ in the
two coexisting phases given by the compositions $x_1$ and $x_2$. That is:

\begin{eqnarray}
\ \mu_A(x_1)=\mu_A(x_2) 
\end{eqnarray}
\begin{eqnarray}
\ \mu_B(x_1)=\mu_B(x_2) 
\end{eqnarray}

\noindent
where (see Fig.~\ref{fig_commontangent}) :

\begin{eqnarray}
\ \mu_A(x)=\Bigl(\frac{\partial g(x)}{\partial x_A}\Bigr)_{p^*,T^f}=g(x)-(1-x)\frac{\partial g(x)}{\partial x} \,\,\,\,\,\
\end{eqnarray}

\begin{eqnarray}
\ \mu_B(x)=\Bigl(\frac{\partial g(x)}{\partial x_B}\Bigr)_{p^*,T^f}=g(x)+x\frac{\partial g(x)}{\partial x} 
\end{eqnarray}
\\
with $x\equiv x_A=1-x_B$. 
\\
\vspace*{0.1cm}

We should remember that Eqs. (24) and (25), as obtained from Eqn.~(23), 
do not only consider the chemical equilibrium of the two species
in the two phases given by $x_1$ and $x_2$ but also assure thermal and
mechanical equilibrium since all the points with which we construct
$g(x)_{T^f,p^*}$ have a common temperature and pressure. Additionally,
it is obvious that for those pressures and temperatures for which
$g(x)_{T^f,p^*}$ is convex ($g''(x)>0$ $\forall x\in[0,1]$) we will have
no phase coexistence. To construct the complete phase diagram $x_A-\rho$
corresponding to a fixed temperature $T^f$ (see Fig.~\ref{fig_werteim})
we repeat the whole process by varying $p^*$.

Finally, tie lines connecting two coexisting points $(\rho_1,x_1)$
and $(\rho_2,x_2)$  (see Fig.~\ref{fig_werteim}) are defined
by all those points $(\rho,x)$  that separate into two phases with
compositions $x_1$ and $x_2$ and total number densities $\rho_1$ and
$\rho_2$ conserving the total number of $A$- and $B$-particles as well
as the total volume. These constraints result in a function (tie line)
$x=x(\rho)$ for any double pair $(\rho_1,x_1)$ and $(\rho_2,x_2)$:

\begin{eqnarray}
\ x(\rho)=\frac{x_1\rho_1(\rho-\rho_2)+x_2\rho_2(\rho_1-\rho)}{\rho_1(\rho-\rho_2)+\rho_2(\rho_1-\rho)} , 
\end{eqnarray}

\noindent            
with domain $\rho\in[\rho_1,\rho_2]$ (here we assume $\rho_1<\rho_2$
without loss of generality) and satisfying:

\begin{eqnarray}
\ x(\rho_i)=x_i \,\,\ ; \,\,\ i\in\{1,2\} \quad .
\end{eqnarray}

\bibliographystyle{apsrev}

\end{document}